\documentclass[%
 reprint,
nofootinbib,
 amsmath,amssymb,
 aps,
 nofootinbib,
]{revtex4-2}

\usepackage{graphicx}
\usepackage{dcolumn}
\usepackage{bm}
\usepackage{booktabs}
\usepackage{color}
\usepackage{hyperref}
\usepackage{booktabs}
\usepackage{umoline}

\usepackage{soul} 

\begin{document}


\title{Asteroseismology using quadrupolar \textit{f}-modes revisited: breaking of universal relationships in the slow hadron-quark conversion scenario}

\author{Ignacio F. Ranea-Sandoval}
\email{iranea@fcaglp.unlp.edu.ar}
 \affiliation{%
 Grupo de Gravitaci\'on, Astrof\'isica y Cosmolog\'ia, Facultad de Ciencias Astron{\'o}micas y Geof{\'i}sicas, Universidad Nacional de La Plata, Paseo del Bosque S/N, 1900, La Plata, Argentina.}
 \affiliation{CONICET, Godoy Cruz 2290, 1425, CABA, Argentina.}
 
\author{Mauro Mariani}
\affiliation{%
 Grupo de Gravitaci\'on, Astrof\'isica y Cosmolog\'ia, Facultad de Ciencias Astron{\'o}micas y Geof{\'i}sicas, Universidad Nacional de La Plata, Paseo del Bosque S/N, 1900, La Plata, Argentina.}
 \affiliation{CONICET, Godoy Cruz 2290, 1425, CABA, Argentina.}

\author{Marcos O. Celi}
\affiliation{%
 Grupo de Gravitaci\'on, Astrof\'isica y Cosmolog\'ia, Facultad de Ciencias Astron{\'o}micas y Geof{\'i}sicas, Universidad Nacional de La Plata, Paseo del Bosque S/N, 1900, La Plata, Argentina.}
 \affiliation{CONICET, Godoy Cruz 2290, 1425, CABA, Argentina.}

\author{M. Camila Rodríguez}
\affiliation{%
 Grupo de Gravitaci\'on, Astrof\'isica y Cosmolog\'ia, Facultad de Ciencias Astron{\'o}micas y Geof{\'i}sicas, Universidad Nacional de La Plata, Paseo del Bosque S/N, 1900, La Plata, Argentina.}
 \affiliation{CONICET, Godoy Cruz 2290, 1425, CABA, Argentina.}

\author{Lucas Tonetto}
\affiliation{%
Dipartimento di Fisica, ``Sapienza'' University of Rome, Piazzale A. Moro, 5. 00185 Roma, Italy}
 \affiliation{INFN, Sezione di Roma, Piazzale A. Moro, 5. 00185 Roma, Italy}

\date{\today}

\begin{abstract}
In this work, we consider polar perturbations and we calculate the frequency and damping time of the quadrupolar fundamental $f$-mode of compact objects, constructed using a wide range of  model-independent hybrid equations of state that include quark matter. We give special attention to the impact of the hadron-quark conversion speed that, in the \textit{slow} case, gives rise to a branch of {\it{slow stable hybrid stars}}. Moreover, we study the validity of universal relationships proposed in the literature and find out that none of them remains valid when slow stable hybrid stars are taken into account. This fact could constrain the applicability of asteroseismology methods with fundamental modes designed to estimate the properties of pulsating compact objects. \  We hope that this result could be tested with the start up of the third-generation gravitational wave observatories, which might shed some light on the $f$-mode emission from compact objects.  
\end{abstract}

\maketitle


\section{Introduction} \label{sec:intro} 

Our understanding of compact stars has undergone a major boost during the last fifteen years with a series of several key observations: the detection of the pulsars PSR J0348+0432 \cite{Antoniadis2013} and PSR J0740+6620 \cite{Cromartie2019,Fonseca:2021rfa} (whose radius has recently been estimated using NICER and XMM-Newton data \citep{riley2021ApJ-j0740,miller2021ApJ-j0740}) established the $2\,M_\odot$ constraint; an alternative constraint obtained only with NICER data can be found in Ref. \cite{salmi:2022tro}; the analysis of the \emph{gravitational waves} (GWs) emitted during the binary neutron star (NS) merger event GW170817 \cite{GW170817-detection} and its electromagnetic counterpart (see, for example, Ref.~\cite{GW170817-em}) constrained the dimensionless tidal deformability $\Lambda$ and the radius of a $\sim 1.4~M_\odot$ NS (see, for example, Refs.~\cite{Raithel2018,Abbott:2018}); finally, using NICER data, two independent estimations of the mass and radius of the isolated pulsar PSR J0030+0451 became available \cite{Riley2019,Miller2019}.

So far, GWs have only been directly observed from compact object mergers (see the modern reviews in Refs. \cite{abbott:2021ppo,abbott:2021gwtc3}), but isolated compact objects are also expected to emit GWs in several astronomical scenarios; one of the most promising situations is the emission of GWs of a proto-NS during a core-collapse supernova \cite{Morozova:2018tgw,TorresForne:2019tao,Radice:2019ctg,vartanyan:2023gso}. The GW emission of isolated NSs is produced by their non-radial oscillations (see, for example, Ref.~\cite{andersson2021Univ} and the references therein). These oscillation modes, known as \emph{quasi-normal modes} (QNMs), have complex frequencies, $\omega=2\pi\nu+i/\tau$, whose real part characterizes the oscillation frequency, $\nu$, and the inverse of its imaginary part, the damping time, $\tau$. In particular, various theoretical works indicate that most of the energy emitted in GWs should be channelled through the (quadrupolar) fundamental $f$-mode  \cite{Morozova:2018tgw,TorresForne:2019tao,Radice:2019ctg,rodriguez:2023taf}. With a network of third-generation GW detectors, errors in the determination of the frequency of QNMs in the kHz band are expected to be of a few tenths of Hz \cite{Pratten:2019sed}. Thus, it is expected that both frequency and damping time of the $f$-mode might be measured by these third-generation detectors \cite{Punturo:2010zz,maggiore:2020scf, Hall:2021gwp}; for this reason, we focus this work on this particular QNM.

The QNMs are sensitive to the equation of state (EoS) used to describe matter inside the NSs. For this reason, understanding the spectrum of QNMs might be a useful tool to extract information about the internal composition of compact objects, unveiling the behavior of matter subject to extreme conditions. Besides this intimate EoS-QNMs relationship, there has been proposed a different approach that considers functions, called \textit{universal relationships} (URs), that relate QNMs and macroscopic quantities in an EoS-independent manner. This method, known as \textit{asteroseismology of NSs} is a promising tool expected to be capable of helping estimate macroscopic quantities of pulsating objects, such as the mass, $M$, or the radius, $R$, after a measurement of the frequency and damping time of a detected mode. Since the seminal paper of Andersson and Kokkotas \cite{Anderson:1998tgw}, many alternative URs have been proposed for the $f$-mode. Such URs include not only versions that contain mass, $M$, and radius, $R$ (see, for example, Refs.~\cite{Benhar:2004gwa,tsui:2005uiq,chirenti:2015PhRvD}) but also moment of inertia, $I$ (see, for example, Refs. \cite{lau:2010ipp,chirenti:2015PhRvD}) and dimensionless tidal deformability, $\Lambda$ \cite{sotani:2021urb}. In addition, URs for other non-radial QNMs have been proposed: $p$-modes \cite{Anderson:1998tgw,Benhar:2004gwa}, $g$-modes associated with sharp hadron-quark phase transitions \cite{Ranea:2018omo,Rodriguez:2021hsw}, axial and polar $wI$-modes \cite{Benhar:2004gwa,tsui:2005uiq,w-modes_universal,ranea-sandoval:2023cmr}.

However, as has been shown in \cite{RaneaSandoval:2022bou}, within certain scenarios, the proposed URs could break and dismantle the application of the asteroseismology method. In particular, in \cite{RaneaSandoval:2022bou}, the authors proved that, in the \textit{slow stable hybrid stars} (SSHS) scenario, the URs previously proposed for the $wI$-modes \cite{Benhar:2004gwa,tsui:2005uiq,w-modes_universal} break for certain combinations of parameters of the phase transition. In particular, this becomes more prominent when the hadron-quark transition pressure is high and close to the respective maximum mass star of the hadronic branch. The SSHSs are a type of hadron-quark hybrid stars (HSs) that arise when the hadron-quark conversion timescale is \textit{slow} compared to the time scale associated with the fundamental radial mode, whose nature marks the (un)stable nature of the compact object (see, for example, Ref. \cite{Pereira:2018pte,LugGrunf-universe:2021} and references therein). Such a theoretical possible scenario has been already explored in other works {using a parametric EoS for the quark sector \cite{Rodriguez:2021hsw,Goncalves:2022ios,RaneaSandoval:2022bou,ranea-sandoval:2023cmr,lugones:2023ama}, the Field Correlator Method EoS \cite{Mariani:2019mhs,Curin:2021hsw, Mariani:2022omh} and different versions of the Nambu-Jona-Lasino model \cite{malfatti:2020dba,Lenzi:2023hsw}. In each work, a common feature is the appearance of} long branches of SSHS after the maximum mass of a given family of compact stars{. In some of these SSHS,} the central energy density can reach values up to a few tens of times the nuclear saturation one. Thus, in this work, we will study if the previously proposed URs for the quadrupolar $f$-mode remain valid when long branches of SSHS that arise when certain conditions are taken into account {(see, Ref. \cite{lugones:2023ama} for more details)}. {In Ref.~\cite{zhao:2022urf}, considering a different scenario, in the context of sharp hadron-quark phase transitions and low-mass \textit{fully stable} HSs\footnote{{We refer as \textit{fully stable} to the stellar configurations that are stable both in the \textit{slow} and \textit{rapid} hadron-quark conversion scenarios, i.e., the stellar configurations up to the maximum mass configuration in the mass mass-radius relationships that have $\partial M/ \partial \varepsilon_c >0$.}}, some of these relationships are revised, and, in Figs. $9$ and $10$ of this work, deviations from universality have been shown to occur.} 
 
Throughout this paper, the radius, $R$, is expressed in km, the mass, $M$, also in km (unless specifically said the contrary), the moment of inertia, $I$, in km$^3$, the frequency, $\nu_f$, in kHz and the damping time, $\tau_f$, in seconds.

The work is structured in the following manner. Section \ref{sec:hybeos} describes aspects of hadron-quark phase transitions and presents the hybrid EoS model used in this paper. Moreover, we introduce some concepts related to the dynamical stability of HSs and SSHS, and present initial results for some HSs attributes, such as the mass, the radius, the moment of inertia and the dimensionless tidal deformability, that would become relevant to explore the validity of $f$-mode URs. In Section \ref{sec:fmode}, we present the basic equations that describe polar modes and explain the numerical scheme used to calculate them, in particular, the $f$-mode. In Section \ref{sec:results}, we show our results for the oscillation frequencies and damping times of the quadrupolar
$f$-mode and the comparison with the existing URs. A summary of the work, a discussion about the astrophysical implications of our results, and our main conclusions are provided in Section \ref{sec:conclus}. {In Appendix \ref{app:tables} we present the coefficients of every relationship analyzed in this work.}

\section{Hybrid Equation of State and SSHS}\label{sec:hybeos} 

\begin{figure*}
    \centering
    \includegraphics[width=0.49\linewidth]{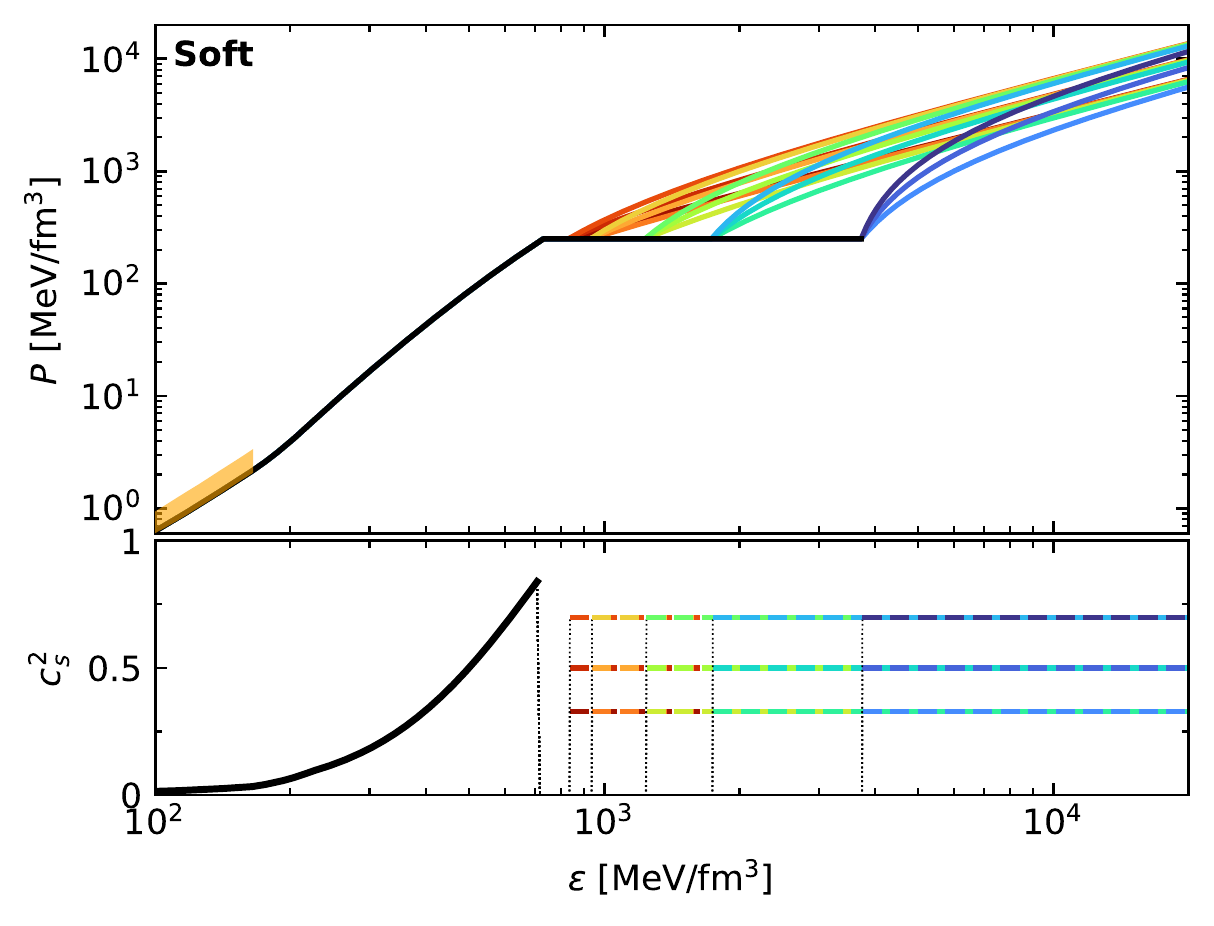}
    \includegraphics[width=0.49\linewidth]{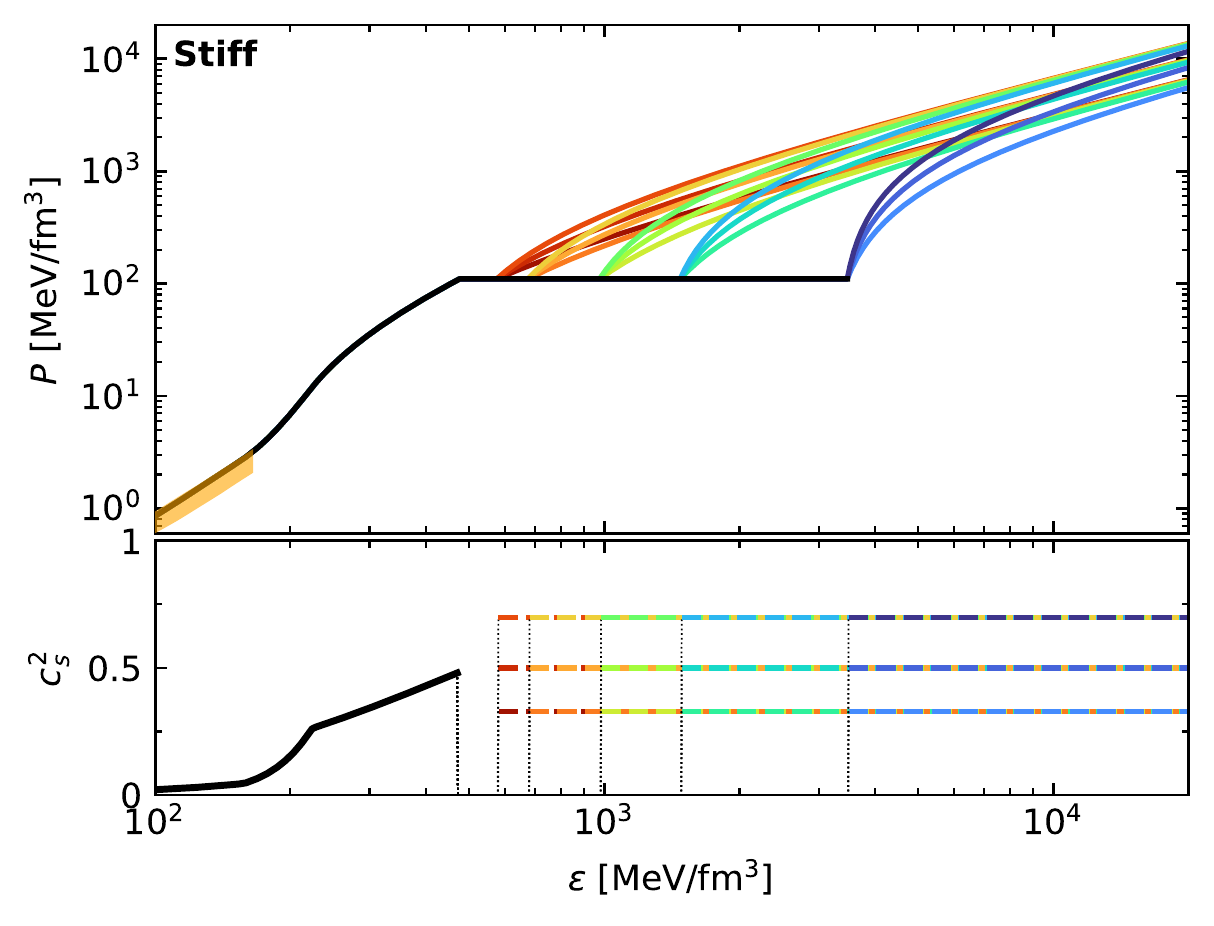}
     \caption{{Pressure-$\varepsilon$ (upper panels) and $c_\mathrm{s}^2$-$\varepsilon$ (lower panels) relationships for the \textit{soft} (left panels) and \textit{stiff} (right panels) cases of the selected hybrid EoS. In black curves, the pure hadron sector; the different colors indicate the hybrid branches for different values for $\Delta \varepsilon$ and $c_\mathrm{s}^2$ (see the main text for details). The orange-colored region in the low-density region of the Pressure-$\varepsilon$ plane indicates the constraint given by chiral EFT up to $1.1\,n_0$ \citep{Hebeler:2013nza, Annala:2020efq}; as can be seen, the \textit{soft} and \textit{stiff} hadronic parametrizations act as limiting cases for this constraint. }}
    \label{fig:EoS}
\end{figure*}

In order to obtain results that are independent of any specific hybrid EoS model, we consider parametric models for both hadron and quark phases, as we have already proposed in previous works \cite{RaneaSandoval:2022bou, ranea-sandoval:2023cmr, lugones:2023ama}. {The crust is described using the GPP Fit to the SLy4 crust given in Ref. \cite{OBoyle:2020peo} (also see Ref. \cite{douchin:2001aue}).} To describe the hadron sector, we use a couple of piecewise polytropic EoSs \cite{OBoyle:2020peo}; these two models (a \textit{soft} and a \textit{stiff} EoS) act as enveloping EoSs to known, microphysically involved, EoS that are consistent with modern astronomical observations. The specific parameters of these two representative polytropic hadronic EoS are given in Table~\ref{tabla:param_selec}. To describe the quark sector in the inner core of HSs, we use the constant speed of sound parametrization (CSS) EoS model \cite{Alford:2013gcf}; this model has three free parameters that characterize the hadron-quark transition pressure, $P_t$, the energy density jump, $\Delta \varepsilon$, and speed of sound, $c_\mathrm{s}$.
In order to consider a wide range of existing quark EoS, we select $15$ combinations of the CSS parameters, that cover different qualitative behaviors for the quark matter and, along with the \textit{soft} and \textit{stiff} hadron EoSs, give rise to different kind of mass-radius relationships and (long) branches of SSHS. For each hadron EoS, we have fixed the value of $P_t$; we have selected $P_t = 250$~MeV/fm$^3$ for the \textit{soft} case and $P_t = 110$~MeV/fm$^3$ for the \textit{stiff} case. We do not study the effect of varying this parameter since if we select low values and demand consistency with the astronomical constraints, there appear long \textit{fully stable} hybrid branches and/or hybrid twin branches (as can be seen in Refs.~\citep{Alvarez-Castillo:2018pve,Jakobus:2021tpo,Goncalves:2022ios}). Hence, we are not interested in these scenarios, since they do not particularly  produce (long) branches of SSHS that allow us to explore the URs rupture hypothesis. On the other hand, taking a larger value for $P_t$ gives rise to maximum mass configurations above 2.3$M_\odot$, which are discarded \cite{shibata:2019cot}. For $\Delta \varepsilon$ and $c_\mathrm{s}$ parameters, we have consider, for both \textit{soft} and \textit{stiff} cases, 
\begin{gather*}
    \Delta \varepsilon = 100,\, 200,\, 500,\, 1000,\, 3000\ \mathrm{MeV/fm^3} \, , \\
    c_\mathrm{s}^2 = 0.33,\, 0.5,\, 0.7 \, .
\end{gather*}
In total, this selection allows us to obtain thirty hybrid EoSs, constructed to qualitatively represent a vast number of existing hybrid EoS in literature and to study, in a model-independent manner, the validity of the URs.

\begin{table}
\centering
\begin{tabular}{cccccccc}
\toprule
 & $\log_{10}\rho_0$ & $\log_{10}\rho_1$ & $\log_{10}\rho_2$  & $\Gamma_1$ & $\Gamma_2$ & $\Gamma_3$ & $\log_{10}K_1$ \vspace{0.1cm}  \\
\toprule
\emph{soft} & 13.902 & 14.45 & 14.58 & 2.752 & 4.5 & 3.5 & -27.22 \\
\emph{stiff} & 13.902 & 14.45 & 14.58 & 2.764 & 8.5 & 3.2 & -27.22 \\
\bottomrule
\end{tabular}
\caption{Parameters of the selected hadronic EoSs constructed with the prescription of Ref.~\cite{OBoyle:2020peo}. {The values of the parameters are selected so that the mass-radius relationships obtained with the resulting EoS act as possible (but not unique) limiting families compatible with the modern astronomical data shown in Fig \ref{fig:MR}.} {Moreover, the two EoSs are also consistent with the borders of the region allowed by  chiral EFT presented in Refs. \citep{Hebeler:2013nza, Annala:2020efq}.}}
\label{tabla:param_selec}
\end{table}

Once selected the parameter sets for the hadron and quark phases of the hybrid EoS, we should model the hadron-quark interface. The nature and main characteristics of the hadron-quark phase transition are essentially determined by the value of the surface tension between the hadronic and quark states of matter, $\sigma _{{\rm hq}}$. There is still no agreement on the numerical value of this physical quantity and a large range of model-dependent values for $\sigma_{\rm hq}$ can be found in the literature (see the recent review on this subject, Ref.~\cite{LugGrunf-universe:2021} and references therein). Despite this uncertainty, there is a general consensus that for low values of \mbox{$\sigma _{{\rm hq}} \sim 5 - 30$~MeV/fm}$^2$, the appearance of a mixed phase, in which hadrons and quarks coexist, is energetically favored (see, for example, Ref. \cite{Alford:2001zr}) and that for larger values of $\sigma _{{\rm hq}} \sim 50 - 300$~MeV/fm$^2$ a sharp hadron-quark phase transition would take place (see, for example, Ref.~\cite{LugGrunf-universe:2021}). 

In the latter case, each phase is constrained to a particular region of the star (quarks confined to the inner core surrounded by an outer core of hadronic matter). This situation leads to EoSs in which the energy density presents a discontinuity at the pressure at which the phase transition occurs. In this scenario, an important aspect that has deep astronomical implications is related to the conversion speed -compared to the characteristic time of radial modes- between hadrons and quarks that are in contact at the interface \cite{Pereira:2018pte}. Although uncertain, there are theoretical arguments indicating that the conversion timescale of the reverse reaction should be \textit{slow}. It is important to recall that, generically speaking, phase transitions are highly collective, non-linear phenomena. This favors the idea that the conversion timescale of the hadron-quark phase transition might not be determined by the process of particles confining or deconfining independently. The general consensus is that a first-order hadron-quark phase transition can take place via nucleation \citep{Olesen:1993ek, Lugones:1997gg,Iida:1998pi,Bombaci:2004mt}. A direct hadronic matter conversion into quark matter in $\beta$ equilibrium is strongly suppressed as it is a high-order weak process. For this reason, nucleation is expected to proceed through an intermediate (flavor-conserving) state (for details, see Refs.~\cite{LugGrunf-universe:2021,Lugones:2015bya}). 

{Despite the fact that the slow conversion regime is a possible physical scenario, recent results presented in Ref. \cite{rau:2022arXivtfp} demonstrated that the branches of SSHS appear even if the hadron-quark phase conversion speed is intermediate. This fact shows that the existence of SSHS occurs in a broader spectrum of possible theoretical scenarios.}

\begin{figure*}
    \centering
    \includegraphics[width=0.7\linewidth]{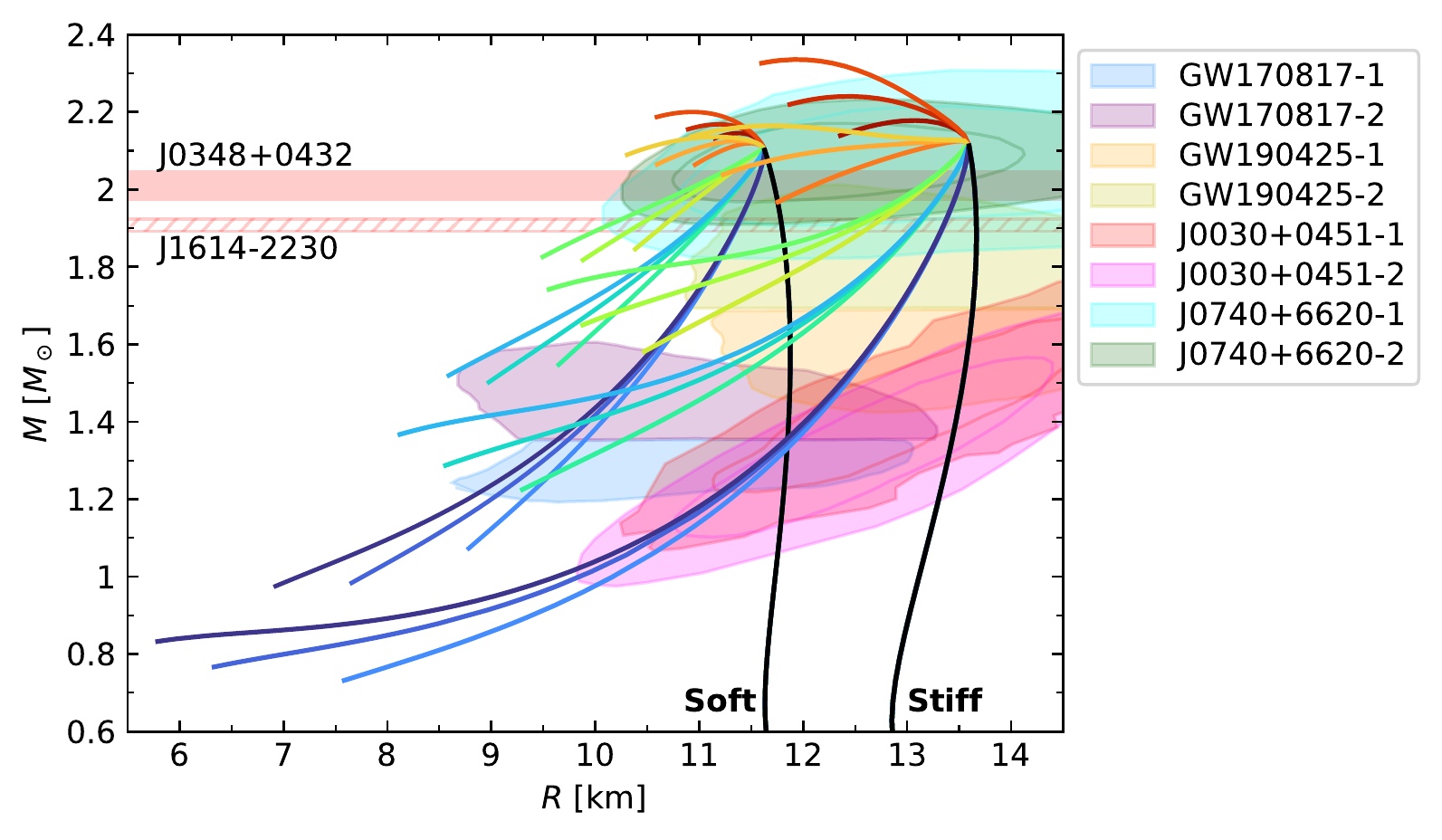}
     \caption{Mass-radius relationships for the selected hybrid EoS. In black curves, the pure hadron branches for \textit{soft} and \textit{stiff} cases; the different colors indicate the hybrid branches for different values for $\Delta \varepsilon$ and $c_\mathrm{s}^2$ (see the main text for details). With colored bars and clouds, we include modern astronomical constraints for compact objects.}
    \label{fig:MR}
\end{figure*}

To sum up, given that both the surface tension $\sigma _{{\rm hq}}$ and the hadron-quark conversion time are not yet fully known, in our work the hadron-quark phase transition is assumed to be sharp and, to focus on the appearance of SSHS, we considered that the conversion speed between phases is \textit{slow}, compared to the typical timescale of radial oscillations. As we already introduce, these assumptions produce branches of SSHS, i.e., possible theoretical compact objects which are stable against linear radial perturbations for which $\partial M/\partial \varepsilon _c < 0$. In this scenario, the mass-radius diagram extends until the \textit{terminal mass} configuration, which is the first stellar configuration for which the fundamental radial mode becomes unstable. {To determine such configurations, the equations that govern linearized radial pulsations of compact stars with the corresponding boundary conditions at the hadron-quark interphase need to be solved (see, Ref. \cite{Pereira:2018pte} for a more detailed discussion on this subject). Such equations (and the corresponding boundary conditions) constitute a Sturm-Liouville problem, for this reason, the first stellar configuration for which the fundamental eigenfrequency becomes imaginary determines the position of the terminal mass in which the SSHS branch ends.} We must emphasize that this configuration might not coincide with the turning points of such curves, in opposition with the \textit{rapid} conversion case, where the terminal mass coincides with the maximum mass configuration. As the SSHS branch is indeed a \textit{twin} branch, it is important to remark that an evolutionary channel in which it might get populated is during the proto-NS stage, as has been discussed in Refs.~\cite{LugGrunf-universe:2021,RaneaSandoval:2022bou}.

{We present the $P$-$\varepsilon$ relationship for the selected hybrid EoSs in the upper panels of Fig.~\ref{fig:EoS}. In addition, as can be seen in these panels, the low-density part of the {\it{soft}} ({\it{stiff}}) limiting EoS is selected to be consistent with the lower (upper) border of the region allowed by chiral EFT presented in Refs. \citep{Hebeler:2013nza, Annala:2020efq}. In the lower part of each panel, we present the squared speed of sound, $c_\mathrm{s}^2$, as a function of the energy density for each of the hybrid EoS constructed for this work. After the hadron-quark phase transition, we see the constant value corresponding to the CSS parametrization EoS used to describe quark matter.}

\begin{figure}
    \centering
    \includegraphics[width=0.99\linewidth]{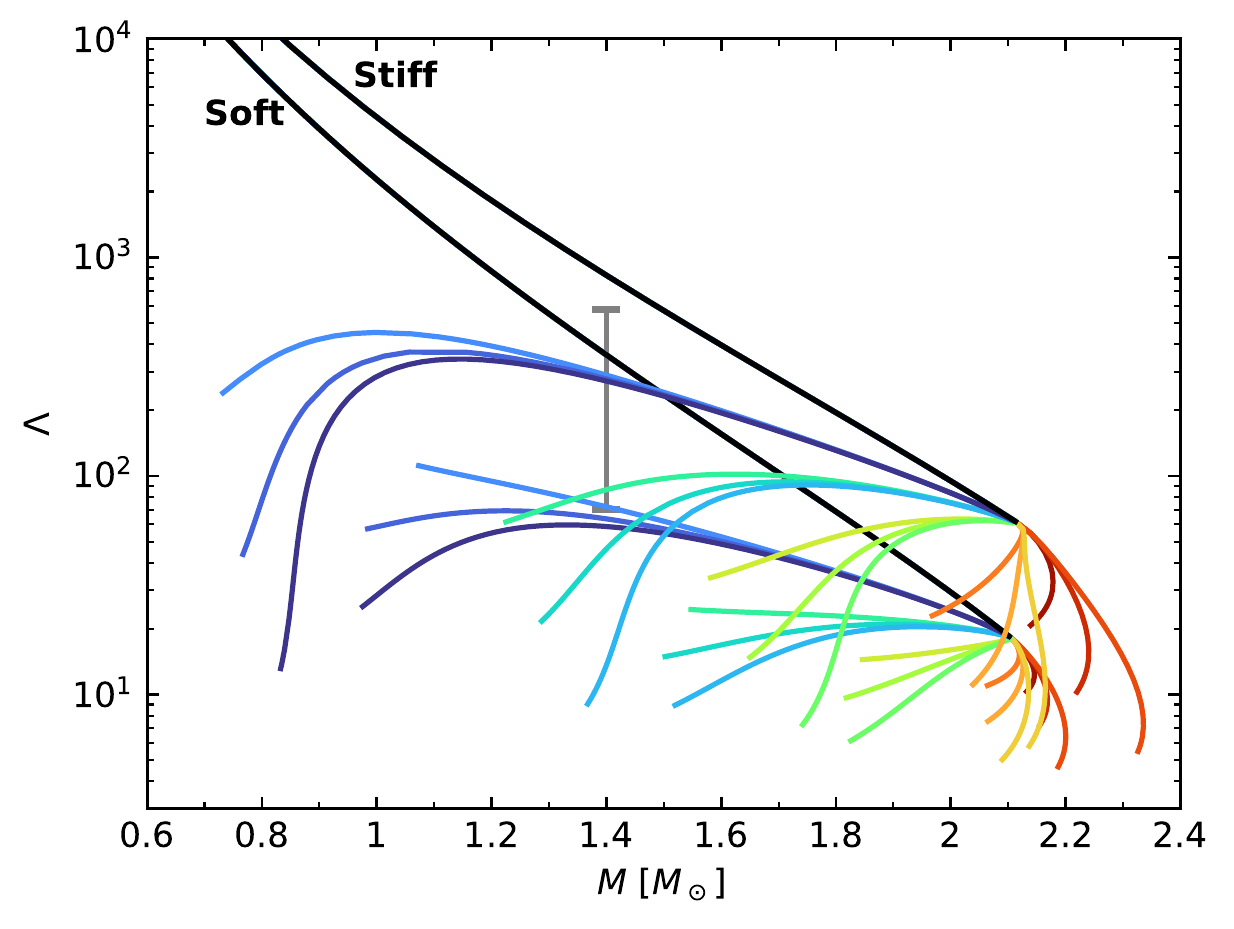}
     \caption{Dimensionless tidal deformability, $\Lambda$, as a function of the mass. In black curves, the pure hadron branches for \textit{soft} and \textit{stiff} cases; the colors used are the same as in Fig.~\ref{fig:MR}. In gray, we show the observational constraint obtained from the analysis of GW170817 and its electromagnetic counterpart \cite{Abbott:2018wiz}.}
    \label{fig:LM}
\end{figure}

\begin{figure}
    \centering
    \includegraphics[width=0.99\linewidth]{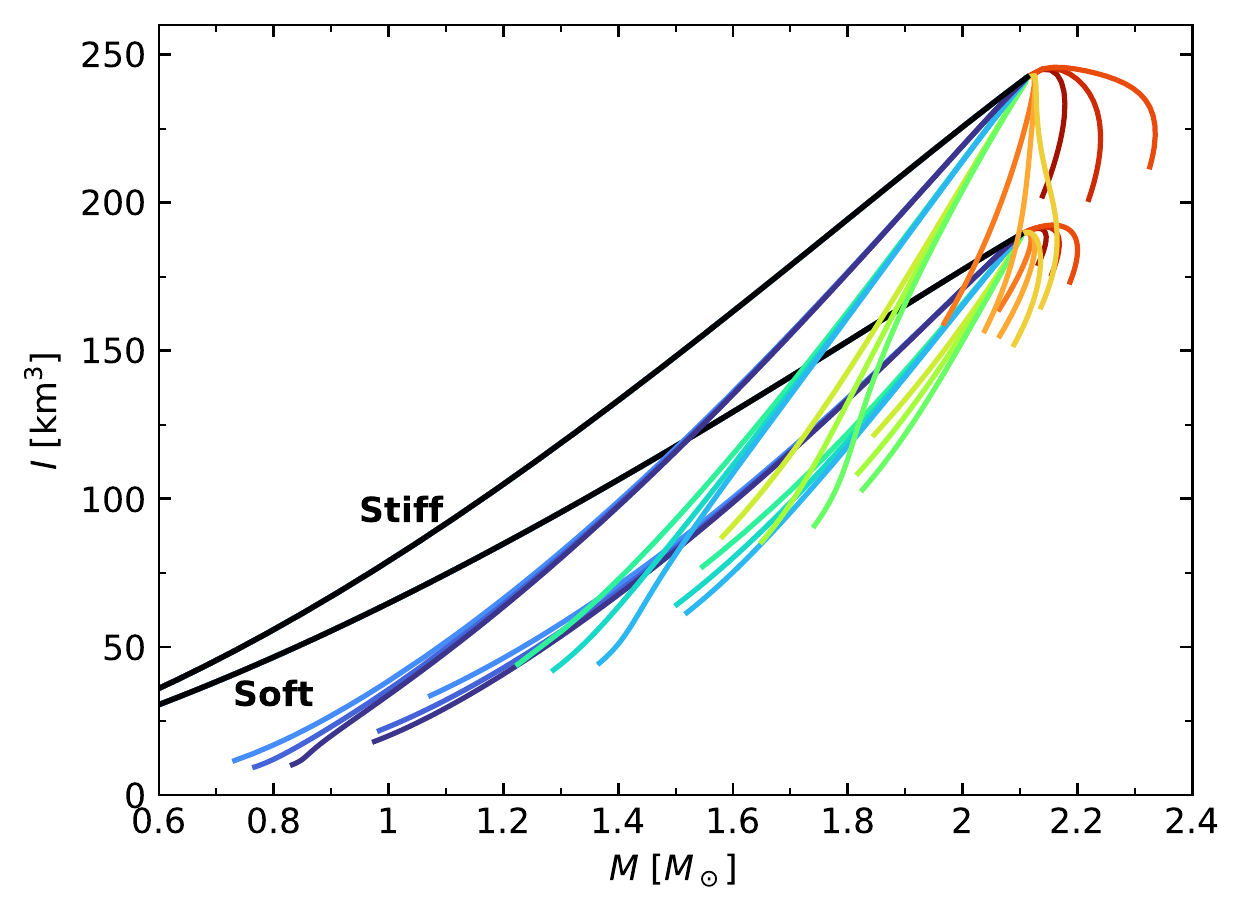}
     \caption{Moment of inertia, $I$, as a function of the mass. In black curves, the pure hadronic branches for \textit{soft} and \textit{stiff} cases; the colors used are the same as in Fig.~\ref{fig:MR}.}
    \label{fig:IM}
\end{figure}

After constructing the hybrid EoS we solved the Tolman-Oppenheimer-Volkoff equations to calculate the different families of compact stars. In addition to the gravitational mass, $M$, and the radius, $R$, of each stellar configuration, we have calculated both the moment of inertia, $I$, and the dimensionless tidal deformability, $\Lambda$. The stability criteria developed in Ref.~\cite{Pereira:2018pte} is implemented to characterize stable and unstable stellar configurations. In Fig.~\ref{fig:MR}, we present the mass-radius relationship for each EoS, together with modern astronomical constraints. Every stellar configuration shown in this plot corresponds to a stable one when a slow hadron-quark phase transition is considered. {Since for the purpose of our work it is necessary to analyze the global behavior of the EoSs and not to look at them separately, we have not labelled them individually. In black we present the pure hadronic configurations, \textit{soft} and \textit{stiff}, and the different color branches correspond to the hybrid configurations with different values of $\Delta \varepsilon$ and $c_\mathrm{s}^2$. Although we have not identified the different EoSs in detail and it is not crucial in order to follow the main results of our work, it is possible to analyze qualitatively the effect of the parameter variation: the colors change from red to blue, corresponding to the variation from small to big values of $\Delta \varepsilon$, respectively; i.e., short branches with bigger masses correspond to smaller values of $\Delta \varepsilon$, and longer branches with smaller masses correspond to bigger values of $\Delta \varepsilon$. Given the same value of $P_t$, the variation of $c_\mathrm{s}$ produces less dramatic effects, obtaining three grouped together branches for $c_\mathrm{s}^2=0.33,0.5,0.7$, with a bigger mass value for the larger value of $c_\mathrm{s}$, e.g., for $P_t=110$~MeV/fm$^3$, it varies from light red ($c_\mathrm{s}^2=0.7$) to dark red ($c_\mathrm{s}^2=0.33$).} In Fig.~\ref{fig:LM}, we show the dimensionless tidal deformability, $\Lambda$, as a function of the mass, together with the restriction obtained after the analysis of the observational data from GW170817 and its electromagnetic counterpart \cite{Abbott:2018wiz}. In Fig.~\ref{fig:IM}, we present the moment of inertia, $I$, as a function of the gravitational mass. It is important to remark that the general behavior for the mass-radius (and $\Lambda$-mass) curve is consistent with those previously reported in Ref.~\cite{RaneaSandoval:2022bou,ranea-sandoval:2023cmr,lugones:2023ama}, showing that stiff EoS, that have been disfavored after the GW170817 merger event, when considering the appearance of long SSHS extended branches, should not be totally discarded yet, and that the merging objects might be SSHSs. In particular, for the stiff case, the families with smaller $\Delta \varepsilon$ values that do not have long enough extended branches do not satisfy the GW170817 constraint; since we are only interested in the study of URs, we do not discard them and keep them in our global model-independent analysis. On the other hand, if we had considered the \textit{rapid} conversion scenario, the configurations would be stable only up to the maximum mass configuration. The appearance of the quark core would destabilize the stellar configurations almost immediately and there would be strong arguments to discard the stiff EoS family.

\section{Non-radial polar perturbations}\label{sec:fmode}

Non-radial oscillations due to polar perturbations can be studied following the seminal works of Lindblom and Detweiler \cite{LD1983,DL1985}, where the main equations needed to be solved are presented. Introducing non-radial perturbations into a spherically symmetric space-time gives rise to a line element that reads:
\begin{eqnarray*}
 {\rm d}s^2 &=& - {\rm e}^{\kappa} (1+r^\ell H_{0 \ell m} Y_{\ell m} e^{i\omega t}) {\rm d}t^2 \\
 && - 2 i\omega r^{\ell+1} H_{1 \ell m} Y_{\ell m} e^{i\omega t} {\rm d}t {\rm d}r \\
&& + {\rm e}^{\lambda} (1 - r^\ell H_{0 \ell m} Y_{\ell m} {\rm e}^{i \omega t}) {\rm d}r^2 \\
         &&+ r^2 (1 - r^\ell K_{\ell m} Y_{\ell m} {\rm e}^{i\omega t}) {\rm d}\Omega^2 \, . \label{pert_ds}
\end{eqnarray*}
Moreover, the components of the Lagrangian perturbation to the fluid are described by:
\begin{eqnarray}
\label{pert_fluid}
\xi^r &=& {\rm e}^{-\lambda /2} r^{\ell -1} W(r)  Y_{\ell m}{\rm e}^{i\omega t} \, , \nonumber \\
\xi^\theta &=& -r^{\ell-2}V(r) \partial_\theta Y_{\ell m}{\rm e}^{i\omega t} \, ,\nonumber \\
\xi^\phi &=& -\frac{r^{\ell -2} V(r)}{\sin^2\theta}  \partial_\phi Y_{\ell m}{\rm e}^{i\omega t} \, , \nonumber
\end{eqnarray}
where {$\omega=2\pi\nu + i/\tau$} is the (complex) frequency {(being $\nu$ the oscillation frequency and $\tau$ the damping time)} of the perturbation, $\kappa$ and $\lambda$ are the functions that describe the background metric, $Y_{\ell m}$ is the spherical harmonics, and $\Omega$ is the solid angle.

In cases where HSs with discontinuous EoS are considered, it is useful to define a variable $X(r)$ (see Ref.~\cite{minuitti2003} for a more detailed discussion), using the following algebraic relationship:
\begin{equation}
    X(r) = \omega ^2 (\varepsilon + P) {\rm e}^{-\frac{\kappa}{2}}V(r) - \frac{{\rm e}^{\frac{\kappa - \lambda}{2}}}{r}P^\prime W(r) + \frac{\varepsilon + P}{2} {\rm e}^{\frac{\kappa}{2}} H_0(r)\,, \nonumber
\end{equation}
where the prime denotes a radial derivative.

Using this approach, inside the star perturbations are fully characterized by a system of fourth-order differential equations for the unknowns $X(r)$, $W(r)$, $K(r)$, and $K_1(r)$ (for simplicity in the presentation, we have eliminated, from the unknowns, the $\ell m$ dependence). These equations are given by:
\begin{eqnarray}
\label{pert_eqs}
 X^{\,'} &=& -\ell r^{-1} X + (\varepsilon+P)e^{\kappa/2} \{ \tfrac{1}{2} (r^{-1}-\kappa'/2) H_0 \nonumber \\
&& + \tfrac{1}{2} [r \omega^2 e^{-\kappa}  + \tfrac{1}{2} \ell(\ell+1)/r] H_1 + \tfrac{1}{2} (3 \kappa'/2 - r^{-1})K  \nonumber \\
&&- \ell(\ell+1) (\kappa'/2) r^{-2} V - r^{-1} [4 \pi (\varepsilon+P) e^{\lambda/2}   \nonumber \\
&& + \omega^2 e^{\lambda/2-\kappa}- r^2 (r^{-2}e^{-\lambda/2} \kappa'/2)'] W \} \, , \nonumber \\
 W^{\,'} &=& -(\ell+1)r^{-1}W + r e^{\lambda/2} [(\gamma P)^{-1} e^{-\kappa/2} X \nonumber \\
&& - \ell(\ell+1)r^{-2} V + \tfrac{1}{2} H_0 + K]\,, \\
K^{\,'} &=& H_0/r + \tfrac{1}{2}\ell(\ell+1)r^{-1}H_1 - [(\ell+1)/r-\kappa'/2]K  \nonumber \\
&& - 8 \pi (\varepsilon+P) e^{\lambda/2} r^{-1} W\,, \nonumber \\
 H_1^{\,'} &=& -r^{-1} [\ell+1+2M e^{\lambda}/r + 4 \pi r^2 e^{\lambda} (P-\varepsilon)] H_1 \nonumber \\
&& + r^{-1} e^{\lambda} [H_0 + K - 16 \pi (\varepsilon+P) V]\,, \nonumber 
\end{eqnarray}
where 
$\gamma $ is the adiabatic index,
\begin{equation}
\label{gamma_pert}
\gamma=\left.\frac{\varepsilon+P}{P}\frac{\Delta P}{\Delta \varepsilon}\right|_{s}\,, \nonumber
\end{equation}
where the variations are performed at a fixed entropy per baryon, $s$\footnote{We are working with cold HSs, so this condition is automatically satisfied.}.

\begin{figure}[t]
    \centering 
    \includegraphics[width=0.99\linewidth]{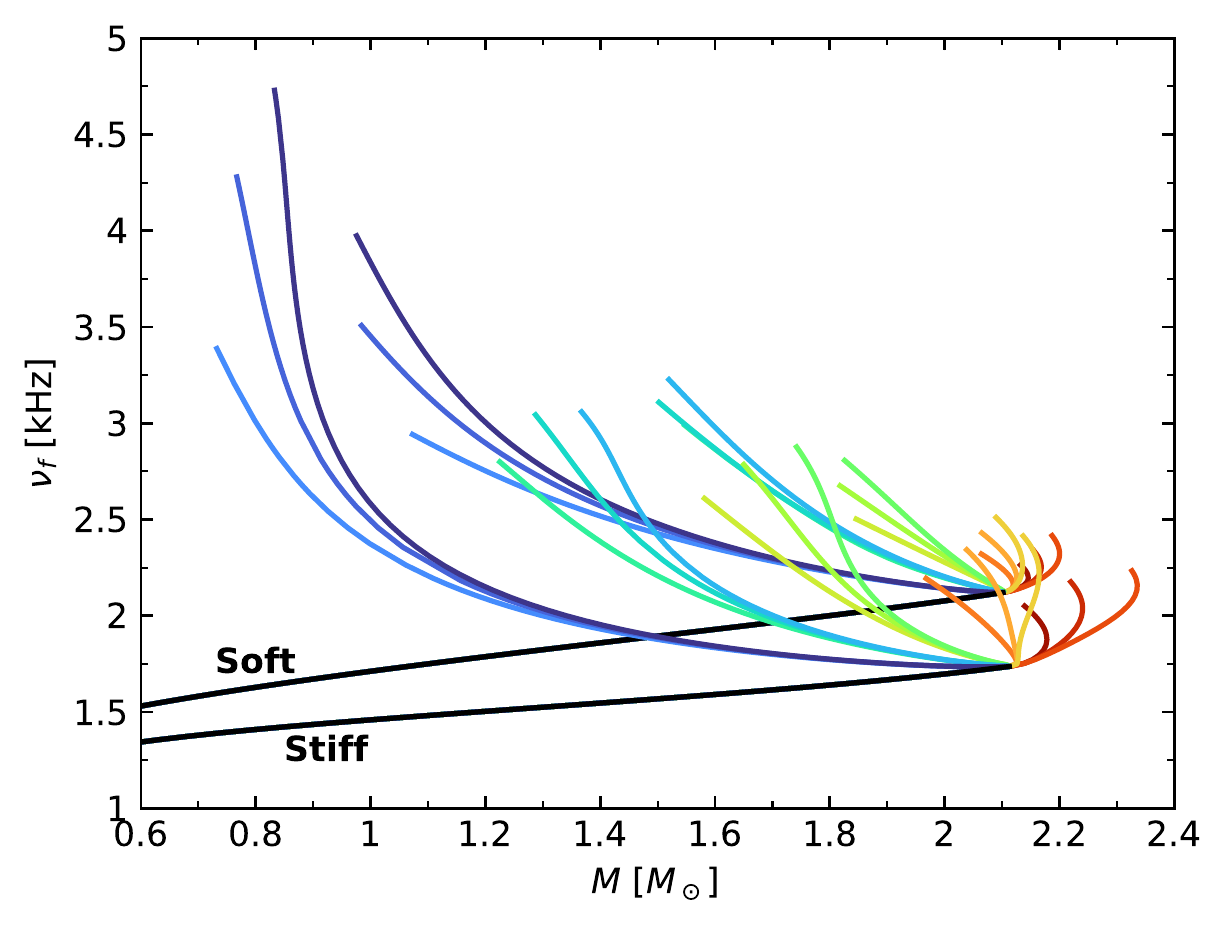}
     \caption{Oscillation frequency of the quadrupolar fundamental $f$-mode, $\nu_f$ as a function of the mass. In black curves, the pure hadron branches for \textit{soft} and \textit{stiff} cases; the colors used are the same as in Fig.~\ref{fig:MR}.}
    \label{fig:nuM}
\end{figure}

\begin{figure}[t]
    \centering
    \includegraphics[width=0.99\linewidth]{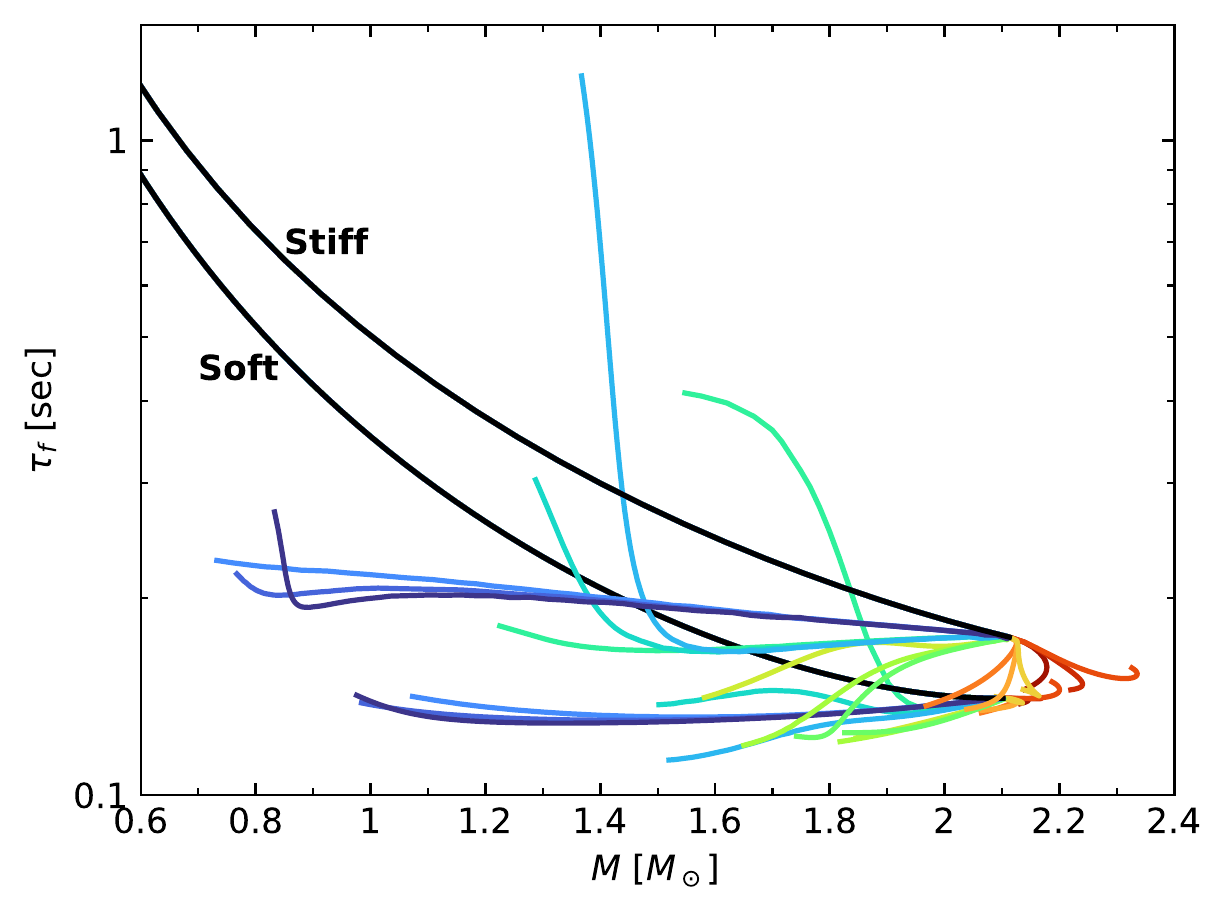}
     \caption{Damping time of the quadrupolar fundamental \mbox{$f$-mode}, $\tau_f$, as a function of the mass. In black curves, the pure hadron branches for \textit{soft} and \textit{stiff} cases; the colors used are the same as in Fig.~\ref{fig:MR}.}
    \label{fig:tauM}
\end{figure}

To deduce the unknown functions $V(r)$ and $H_0(r)$ algebraic expressions can be used. We focus on the quadrupolar perturbations ($\ell=2$) because they are expected to dominate the emission of GWs.

Outside the star, the perturbation equations reduce to the Zerilli second-order differential equation \cite{zerilli1970,fackerell1971,chandrasekhardetweiler1975}. The numerical values of the complex quasi-normal modes, $\omega=2\pi\nu+i/\tau$, are obtained by imposing, to the Zerilli function, a purely outgoing wave behavior at infinity (for details related to the numerical method, see, for example, Ref.~\cite{Tonetto:2020dgm}, and references therein).

We have calculated the quadrupolar $f$-mode for all stable compact objects constructed using the thirty hybrid EoSs presented in Section~\ref{sec:hybeos}. In Figs.~\ref{fig:nuM} and \ref{fig:tauM}, we present, as a function of the mass, the frequency, $\nu_f$, and the damping time of this mode, respectively. We can see that, independently of the hybrid EoS, in the cases in which twin SSHS exist, they have a monotonic increase with the central density and always have larger frequencies compared to their hadronic siblings. In particular, for the cases for which the discontinuity of the energy density is $\Delta \varepsilon = 3000$~MeV/fm$^3$, the frequency of the fundamental mode might be larger than $3$~kHz for stellar configurations near the terminal one. On the other hand, as can be seen from Fig.~\ref{fig:tauM}, when $\tau_f$ is analyzed, the situation has not always a monotonic behavior with the central density and it is strongly model dependent; SSHSs have in general shorter damping times, with \mbox{$\tau_f^{\rm{SSHS}} \sim 0.1-0.2$}~sec, compared to the hadronic twin configuration, but for some particular cases, $\Delta \varepsilon = 1000$~MeV/fm$^3$, there exist configurations with largely greater damping times. 

\begin{figure*}[t]
    \centering
    \includegraphics[width=0.45\linewidth]{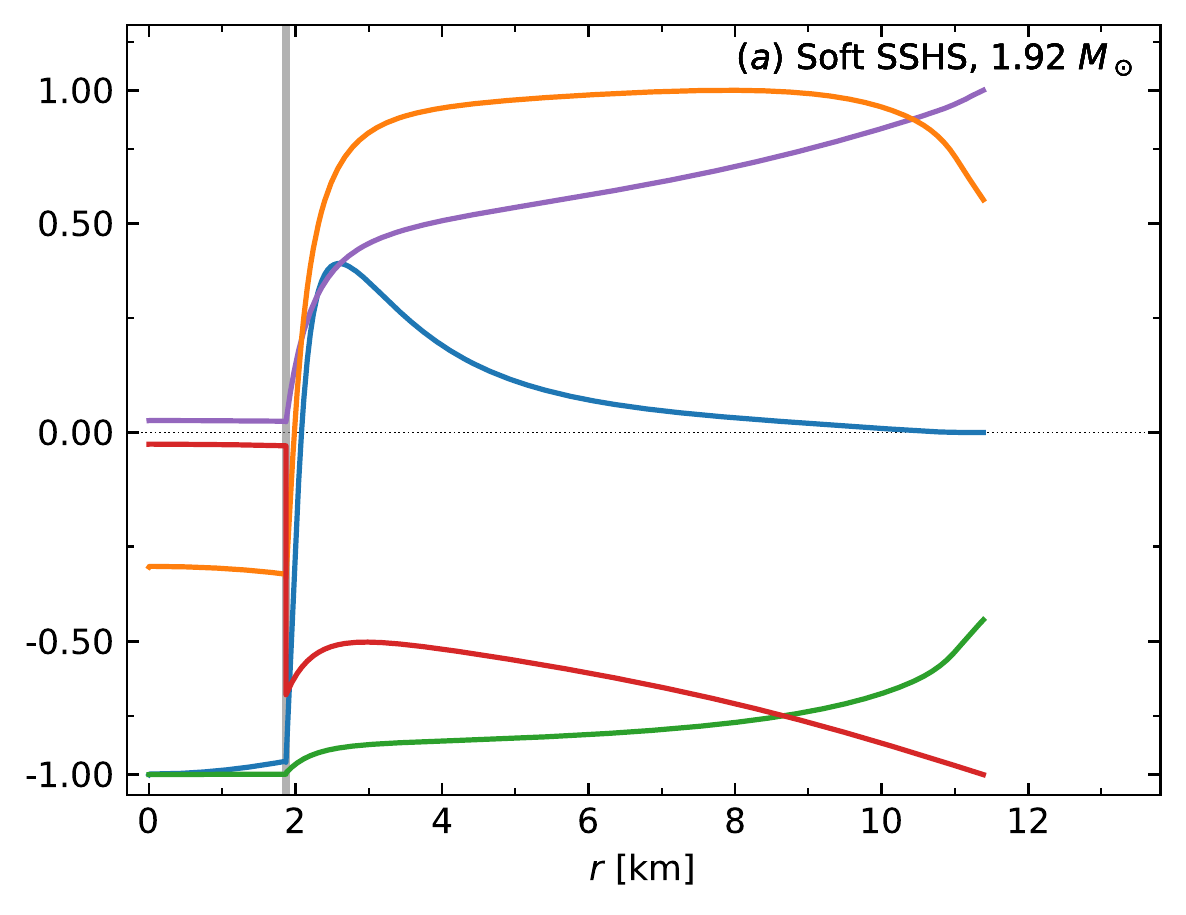}
    \includegraphics[width=0.45\linewidth]{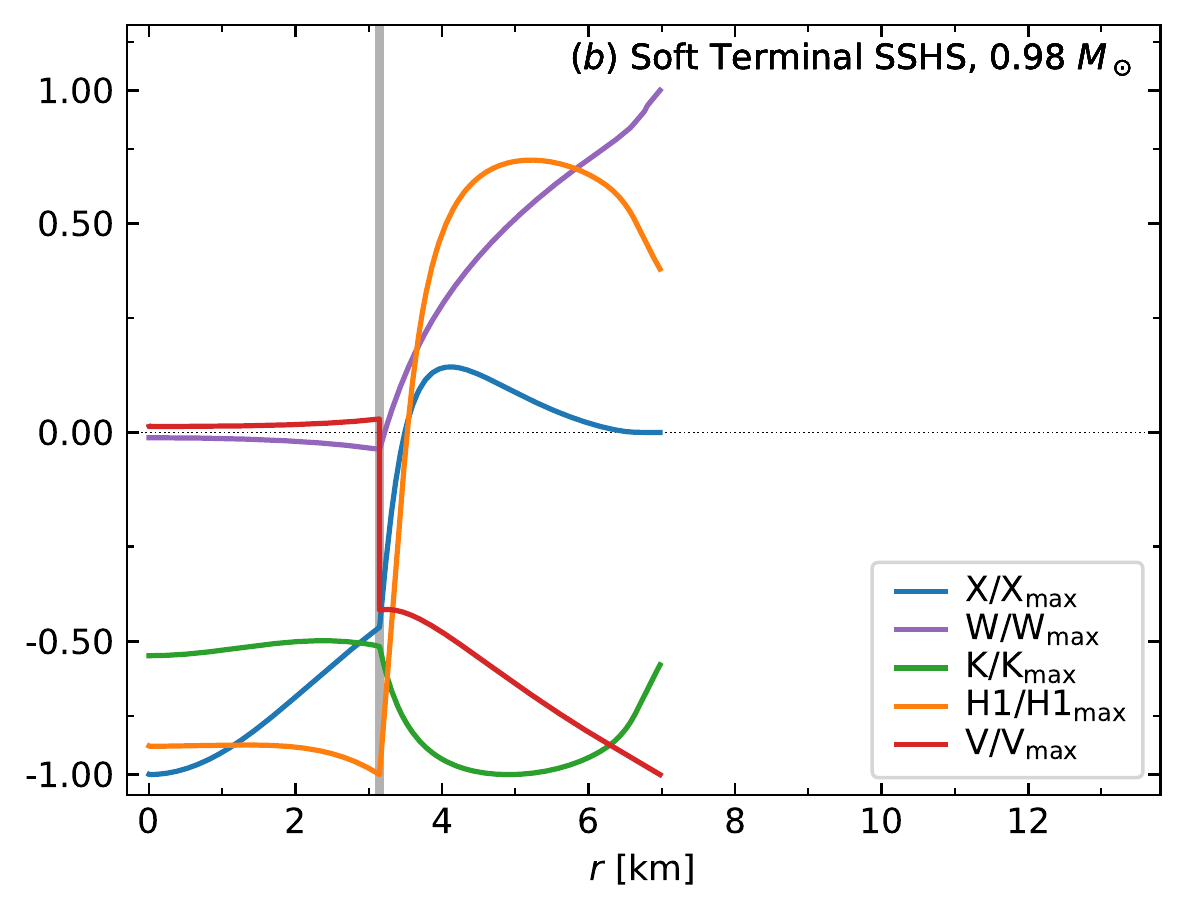}
    \includegraphics[width=0.45\linewidth]{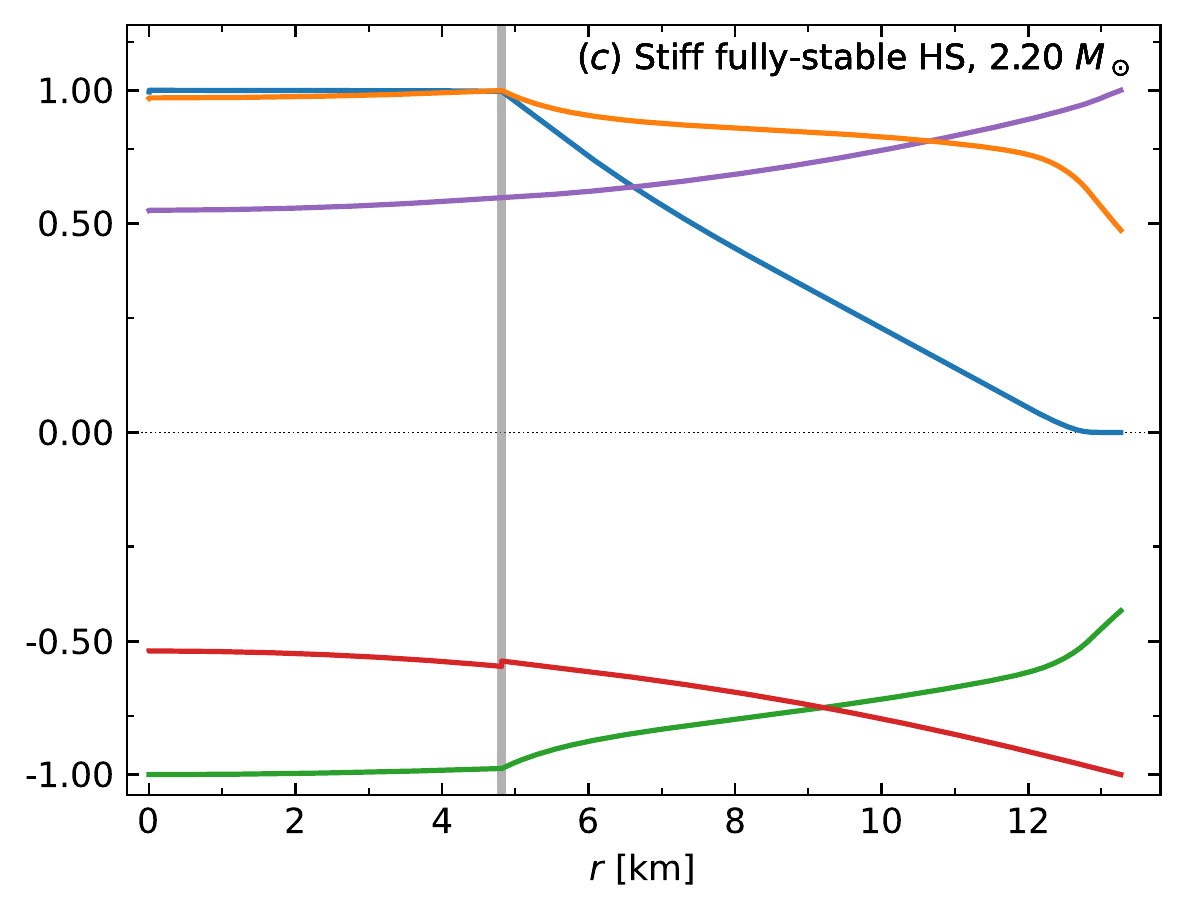}
    \includegraphics[width=0.45\linewidth]{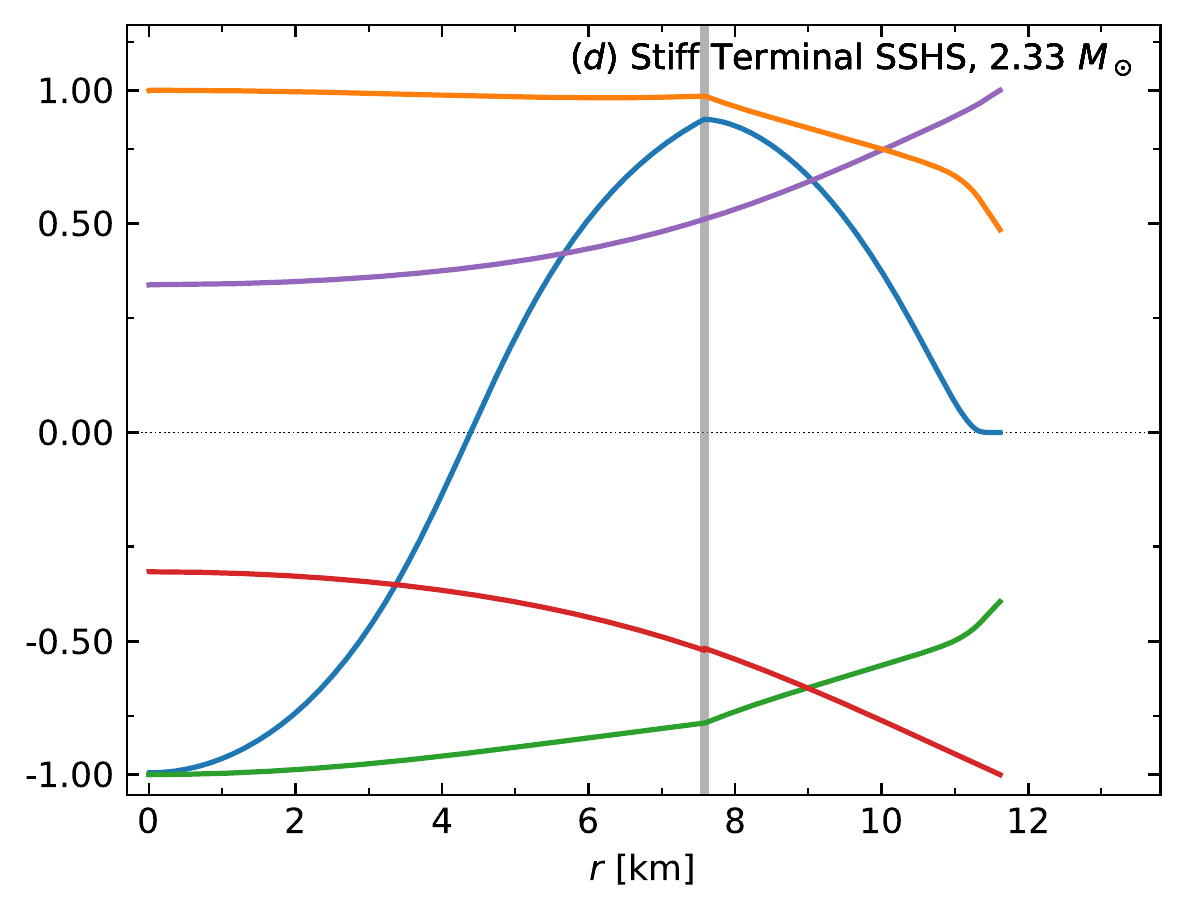}    
    \caption{{Normalized eigenfunctions for different compact objects constructed using the {\textit{soft}, $\Delta \epsilon=3000$~MeV/fm$^3$, $c_s^2=0.70$ EoS -panels $(a)$ and $(b)$-, and the \textit{stiff}, $\Delta \epsilon=100$~MeV/fm$^3$, $c_s^2=0.70$ EoS -panels $(c)$ and $(d)$-}. The gray vertical line represents the radial coordinate in which the hadron-quark phase transition occurs. Left panels present high mass HSs, corresponding to configurations near to the beginning of the hybrid branch; for the \textit{soft} case -panel $(a)$- this configuration corresponds to a high mass SSHS, while for the \textit{stiff} case -panel $(c)$- this configuration corresponds to a fully stable HS, since for this EoS the hybrid branch begins before the maximum mass configuration. The right panels present the terminal SSHS configurations for each EoS. For clarity in the node identification, the ordinate axis is presented in an inverse hyperbolic sine scale, that is approximately linear near the origin and becomes logarithmic for larger positive or negative values. The 1-node family of fundamental modes is seen to appear as some of the fluid and metric eigenfunctions gain a node.}}
    \label{fig:autofunc}
\end{figure*}

{In Fig.~\ref{fig:autofunc}, we present the perturbing functions for four particular stellar configurations constructed using the \textit{soft}, $\Delta \epsilon=3000$~MeV/fm$^3$, $c_s^2=0.70$ EoS -panels $(a)$ and $(b)$-, and the \textit{stiff}, $\Delta \epsilon=100$~MeV/fm$^3$, $c_s^2=0.70$ EoS -panels $(c)$ and $(d)$-. These two EoSs were selected since one, the \textit{soft} one, presents a long extended stability branch with low mass SSHS, and the other, the \textit{stiff} one, presents a short extended stability branch but a high population of fully stable HSs before the maximum mass configuration. For both EoS we present the results for a high mass HS at the beginning of the hybrid branch -panels $(a)$ and $(b)$-, and the results for the terminal SSHS -panels $(b)$ and $(d)$-. As can be seen from  this figure, the perturbing functions of the $f$-mode of the SSHSs shown present 1-node as described in  Ref.~\cite{zhao:2022urf}. Despite this, the behavior of the perturbation functions can not be classified as either the 1-node I or the 1-node II families of such work. In panel $(a)$, for the \textit{soft} high mass SSHS case, only the $H_1$ and $X$ functions have a node. For the same EoS, the situation changes for the low-mass terminal SSHS, panel $(b)$,  where we can see that also the $W$ function gains a node. The case presented in panel $(c)$ of Fig.~\ref{fig:autofunc} shows the \textit{stiff} fully stable HS case in which the $f$-mode eigenfunctions do not have any nodes (see the bottom panels of Fig. 11 in Ref. \cite{zhao:2022urf}). The situation changes for the \textit{stiff} terminal mass SSHS, as the $X$ function gains a node.}

\section{\label{sec:results} Breaking of Universal Relationships for the fundamental mode}

As stated in Section \ref{sec:intro}, several URs have been proposed for different non-radial oscillation modes of compact objects. Some of these relationships have been argued to be useful, after a measurement of the frequency and damping time of a mode, to determine macroscopic quantities of the vibrating stellar configuration. We focus our attention on those related to the quadrupolar $(\ell =2)$ $f$-mode, $\omega_f = 2\pi \nu_f + i/\tau_f$, with $\nu_f$ the frequency and $\tau_f$ the damping time of this particular mode.

The first UR for $f$-modes was proposed in Ref.~\cite{Anderson:1998tgw}, where the idea of NS asteroseismology was coined. The Andersson Kokkotas (AK) fits, which involve the mass and radius of the objects, were revised with modern EoSs in the works by Benhar, Ferrari and Gualtieri (BFG) \cite{Benhar:2004gwa} and by Chirenti, de Souza and Kastaun (CSK) \cite{chirenti:2015PhRvD}, among others. These URs read:
\begin{gather}
    \nu _ f = a_\nu +b_\nu \left(\frac{M}{R^3}\right)^\frac{1}{2} \, , \label{eq:freqfit} \\
    \frac{R^4}{M^3 \tau_f} = a_\tau + b_\tau \frac{M}{R} \, , \label{eq:taufit}
\end{gather}
where $M$ and $R$ are the mass and radius, the coefficients $a_\nu$, $b_\nu$ from those previous works are shown in Table~\ref{table:fitsnu}, and $a_\tau$ and $b_\tau$, in Table~\ref{table:fitstau} {of Appendix \ref{app:tables}}. We present, in Fig.~\ref{fig:uniMRnu}, the results related to $\nu _f$ for the UR of Eq.~\ref{eq:freqfit}. There it can be noticed that the hadronic branches have the same universal behavior studied in previous works, but can be also noticed a clear deviation from universality that the presence of the  extreme SSHSs produced. Despite the large spread of the results, the loss of universality is more evident for cases in which $\Delta \varepsilon \gtrsim 2000$ MeV/fm$^3$. A similar situation is presented in Fig.~\ref{fig:uniMRtau} for the damping time; the traditional stable branches follow the proposed URs, while the extended stability branch clearly break these relationships. Both in these figures and in all the following ones, it is important to remind that, although we separate the mass-radius curves in two segments, one black and one in color, according to the absence (or presence) of quark matter in the inner core, not all of the color configurations are SSHSs. Only the color configurations after the maximum mass one corresponds to SSHSs, and so, it is not expected that the curves corresponding to smaller values of $\Delta \varepsilon$ -the reddish and orange ones-, that have most of its hybrid (colored) branches populated by \textit{fully stable} configurations, deviate much from the URs.

\begin{figure}[t]
    \centering
    \includegraphics[width=0.99\linewidth]{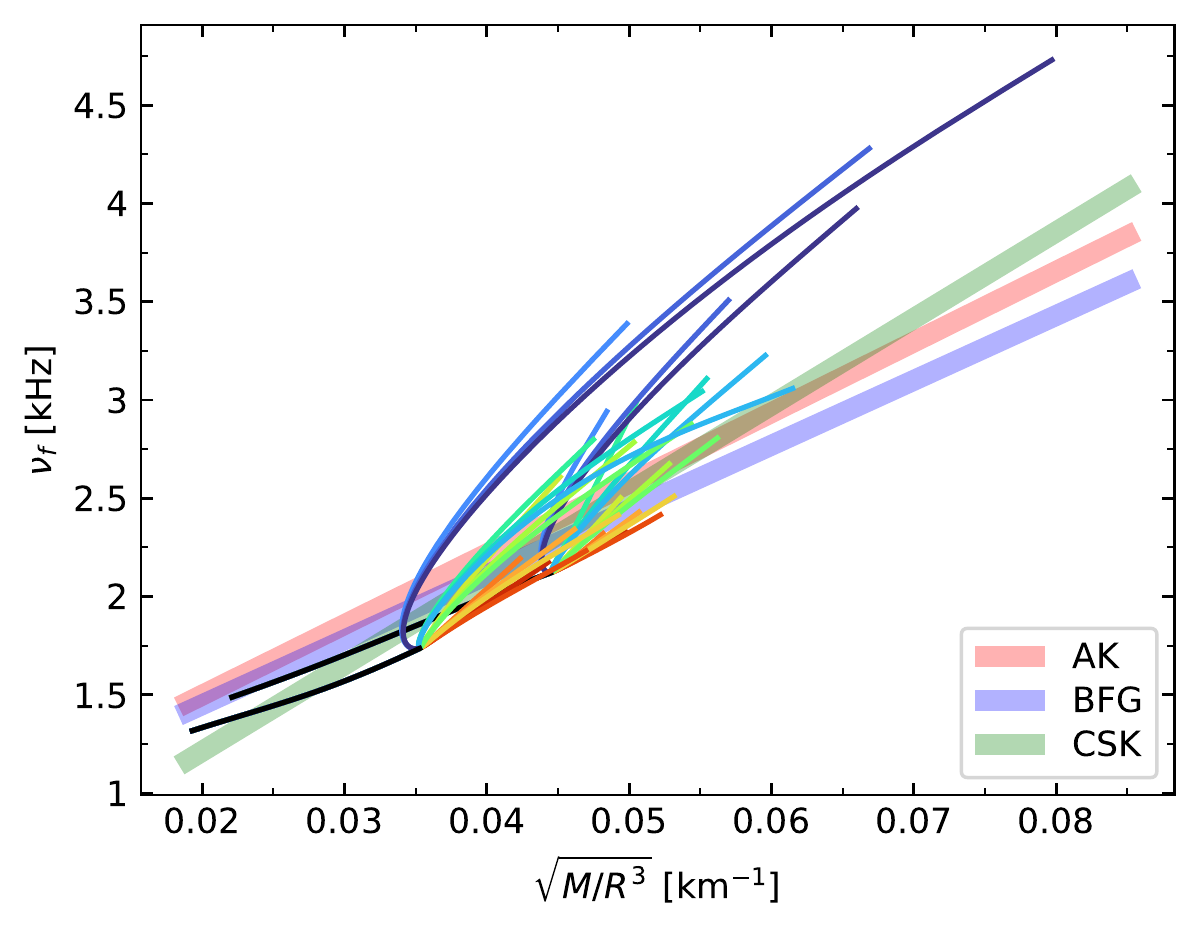}
     \caption{Breaking of URs for the frequency of the quadrupolar fundamental $f$-mode presented in Refs.~\cite{Anderson:1998tgw,Benhar:2004gwa,chirenti:2015PhRvD}. In black curves, the pure hadron branches for \textit{soft} and \textit{stiff} cases; the colors used are the same as in Fig.~\ref{fig:MR}. We show the relationships proposed in the literature and the clear deviation from universality produced by SSHSs presence.}
    \label{fig:uniMRnu}
\end{figure}

\begin{figure}[t]
    \centering
    \includegraphics[width=0.99\linewidth]{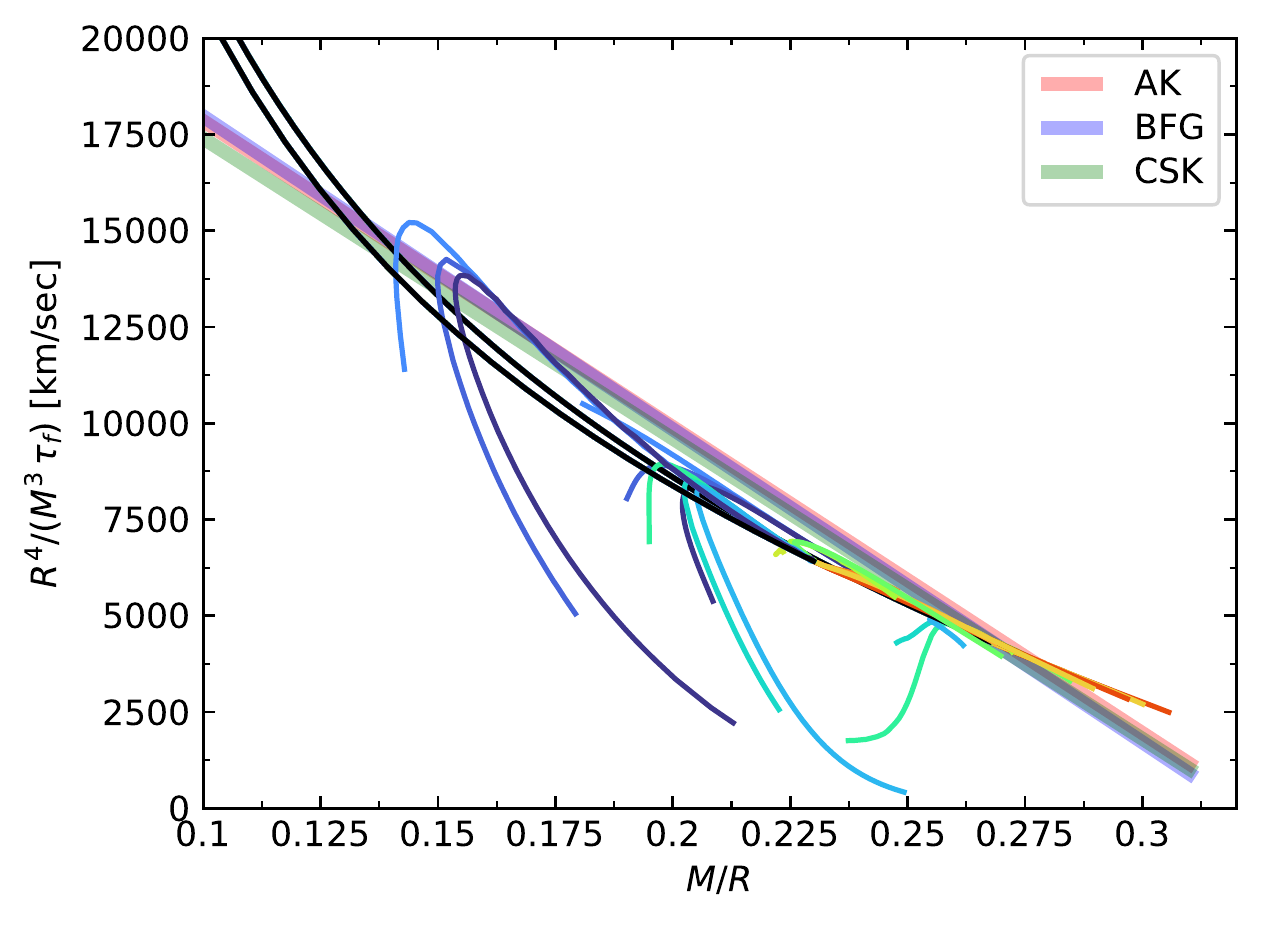}
     \caption{Breaking of URs for the damping time of the quadrupolar fundamental $f$-mode presented in Refs.~\cite{Anderson:1998tgw,Benhar:2004gwa,chirenti:2015PhRvD}. In black curves, the pure hadron branches for \textit{soft} and \textit{stiff} cases; the colors used are the same as in Fig.~\ref{fig:MR}. We show the relationships proposed in the literature and the clear deviation from universality produced by SSHSs presence.}
    \label{fig:uniMRtau}
\end{figure}

\begin{figure}[t]
    \centering
    \includegraphics[width=0.99\linewidth]{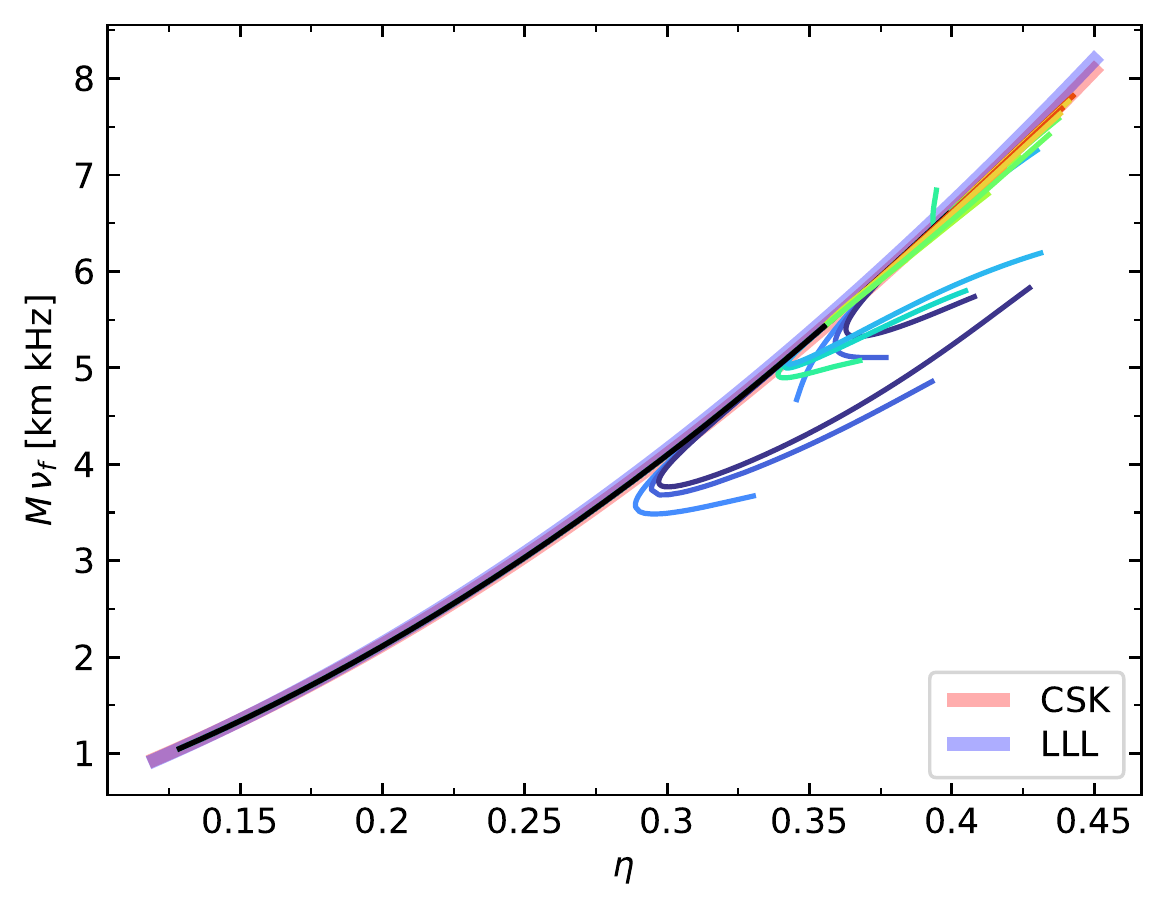}
     \caption{Breaking of URs that include the moment of inertia $I$ for the frequency of the quadrupolar fundamental $f$-mode presented in Refs.~\cite{lau:2010ipp,chirenti:2015PhRvD}. In black curves, the pure hadron branches for \textit{soft} and \textit{stiff} cases; the colors used are the same as in Fig.~\ref{fig:MR}. We show the relationships proposed in the literature in colored wide curves and the clear deviation from universality produced by SSHSs presence.}
    \label{fig:uniinertianu}
\end{figure}

\begin{figure}[t]
    \centering
    \includegraphics[width=0.99\linewidth]{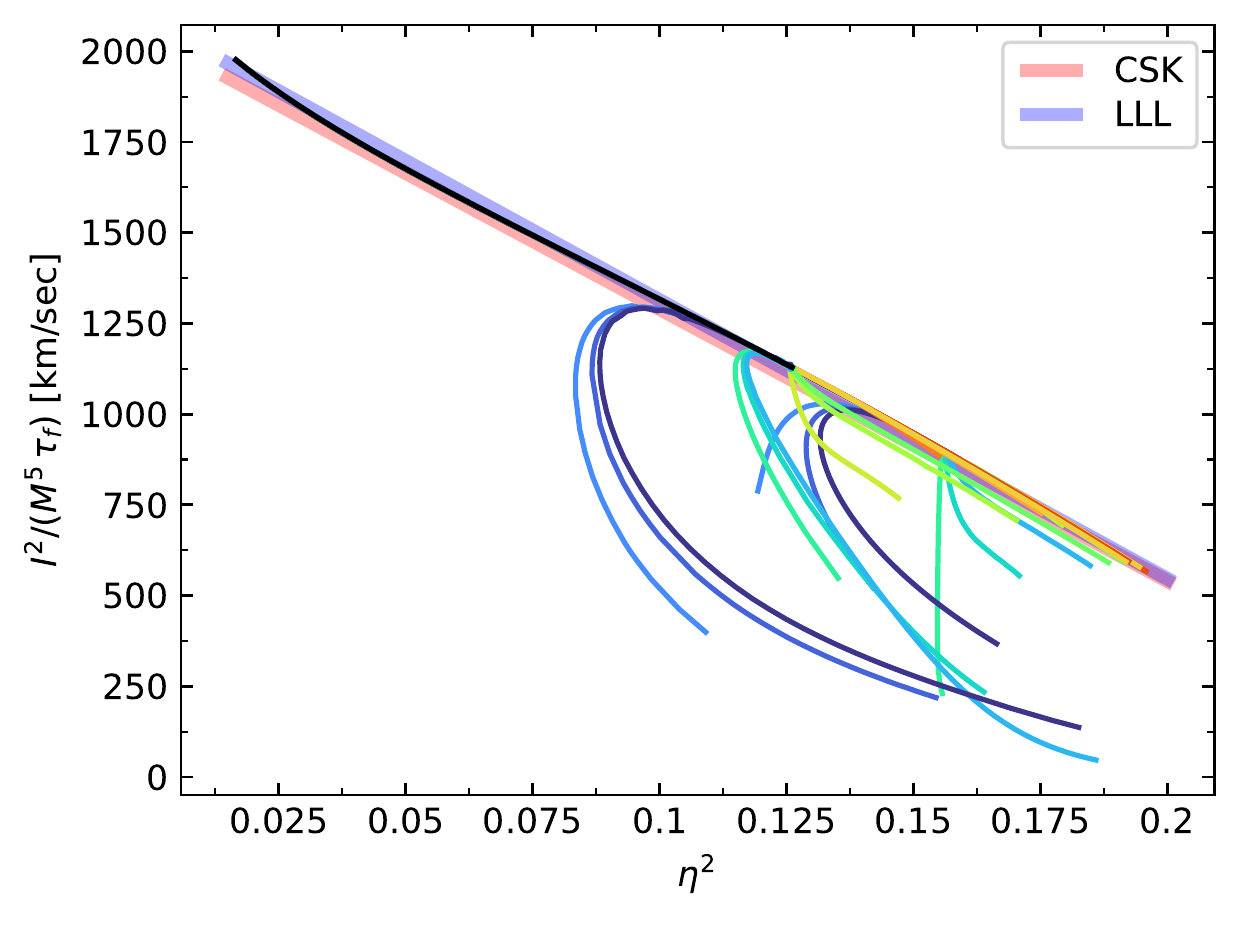}
     \caption{Breaking of URs that include the moment of inertia $I$ for the damping time of the quadrupolar fundamental $f$-mode presented in Refs.~\cite{lau:2010ipp,chirenti:2015PhRvD}. In black curves, the pure hadron branches for \textit{soft} and \textit{stiff} cases; the colors used are the same as in Fig.~\ref{fig:MR}. We show the relationships proposed in the literature and the clear deviation from universality produced by SSHSs presence.}
    \label{fig:uniinertiatau}
\end{figure}

More recently, URs including the mass and the moment of inertia, $I$, have been found in the work by Lau, Leung and Lin (LLL) \cite{lau:2010ipp} (and revised, for example, in the work by CSK \cite{chirenti:2015PhRvD}). The proposed URs read:
\begin{gather}
    M \nu _ f = a^I_\nu +b^I_\nu \eta + c^I_\nu \eta ^2 \, , \label{eq:inertianu} \\
    \frac{I^2}{M^5 \tau_f} = a^I_\tau + b^I_\tau \eta ^2 \, , \label{eq:inertiatau}
\end{gather}
where $\eta=\sqrt{M^3/I}$ is the effective compactness. Coefficients for the fits from these previous works are presented in Table~\ref{table:inertianu} and \ref{table:inertiatau}\footnote{In Refs.~\cite{lau:2010ipp,chirenti:2015PhRvD}, note the factor $2\pi$ relating $\omega_R$ and the frequency $\nu_f$ that we use in this paper.} {of Appendix \ref{app:tables}}. In Fig.~\ref{fig:uniinertianu}, we present the URs of Refs.~\cite{chirenti:2015PhRvD,lau:2010ipp}, that involve $I$ and $M$, for the frequency $\nu _f$ and show the clear deviation from universality that the appearance of the SSHSs generate. As this relationship has been shown to be tighter, deviation from universality is less dramatic and only occurs for really high values of $\Delta \varepsilon\gtrsim 1000$~MeV/fm$^3$. A similar situation is presented in Fig.~\ref{fig:uniinertiatau} for the damping time, but in this case, universality is more notoriously lost, from values of $\Delta \varepsilon \gtrsim 500$~MeV/fm$^3$.

\begin{figure}[t]
    \centering
    \includegraphics[width=0.99\linewidth]{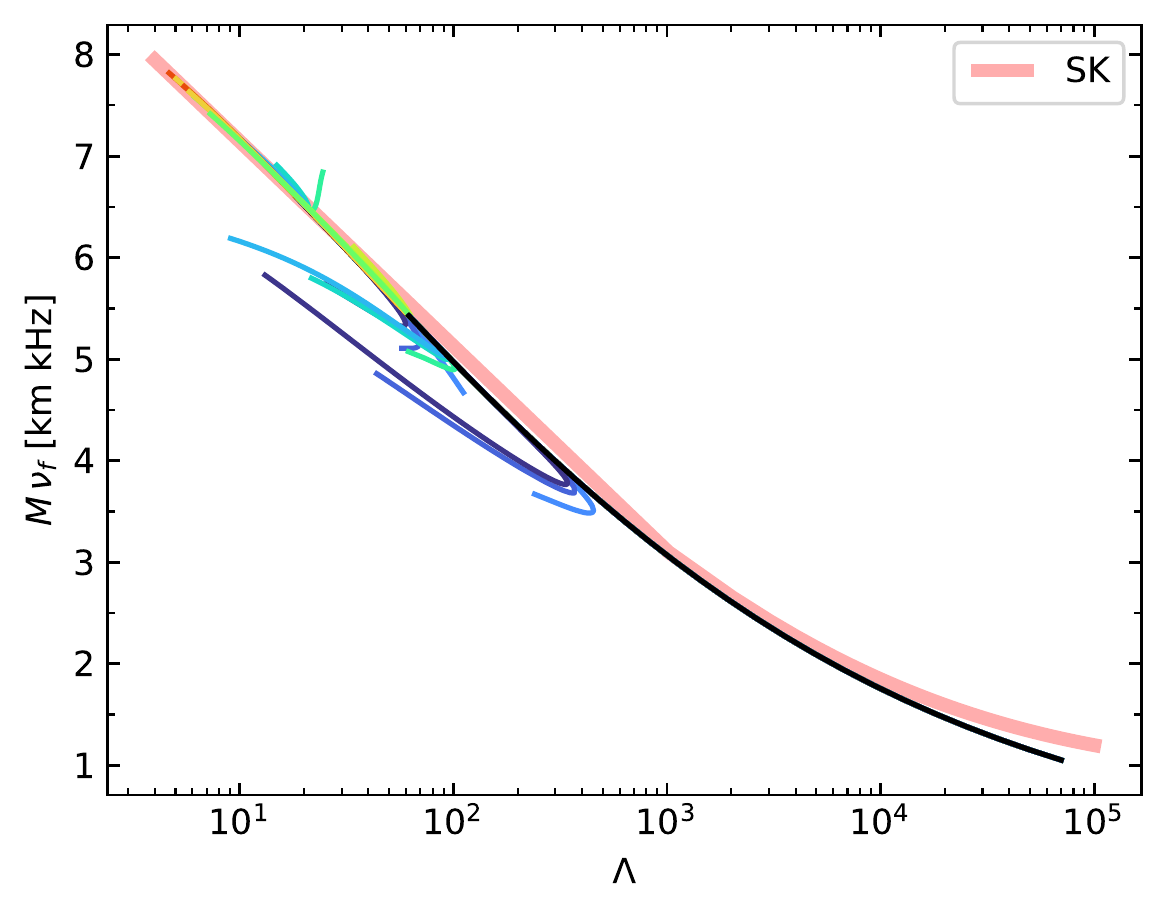}
     \caption{Breaking of URs that include the dimensionless tidal deformability $\Lambda$ for the frequency of the quadrupolar fundamental mode presented in Ref.~\cite{sotani:2021urb}. In black curves, the pure hadron branches for \textit{soft} and \textit{stiff} cases; the colors used are the same as in Fig.~\ref{fig:MR}. We show the relationship proposed in the literature and the clear deviation from universality produced by SSHSs presence.}
    \label{fig:unilambdanu}
\end{figure}

\begin{figure}[t]
    \centering
    \includegraphics[width=0.99\linewidth]{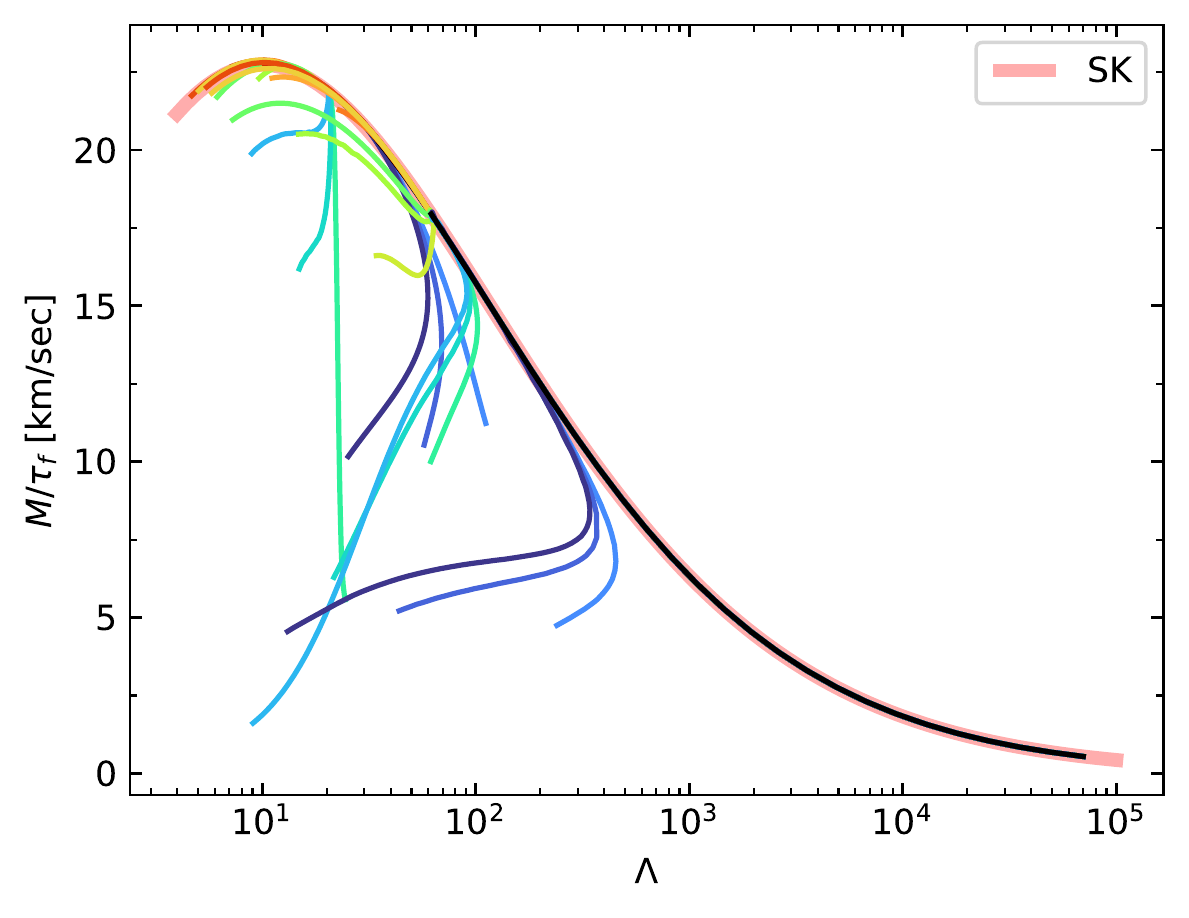}
     \caption{Breaking of URs that include the dimensionless tidal deformability $\Lambda$ for the damping time of the quadrupolar fundamental mode presented in Ref.~\cite{sotani:2021urb}. In black curves, the pure hadron branches for \textit{soft} and \textit{stiff} cases; the colors used are the same as in Fig.~\ref{fig:MR}. We show the relationship proposed in the literature and the clear deviation from universality produced by SSHSs presence.}
    \label{fig:unilambdatau}
\end{figure}

Finally, the last URs that we analyze are those proposed by Sotani and Kumar (SK) that include $M$ and $\Lambda$ and are proposed in Ref. \cite{sotani:2021urb}:
\begin{gather}
    M\nu_f = a^\Lambda_\nu + b^\Lambda_\nu x +  c^\Lambda_\nu x^2 + d^\Lambda_\nu x^3 + e^\Lambda_\nu x^4 + f^\Lambda_\nu x^5 \, ,  \label{eq:nudimL} \\
    M/\tau_f = 10^{f^\Lambda_\tau(x)} \, ,  \label{eq:taudimL}
 \end{gather}
 where 
 \begin{gather} \label{taudimL2}
 f^\Lambda_\tau(x) = a^\Lambda_\tau + b^\Lambda_\tau x +  c^\Lambda_\tau x^2 + d^\Lambda_\tau x^3 + e^\Lambda_\tau x^4 + f^\Lambda_\tau x^5 \, ,
 \end{gather}
with $x = \log (\Lambda)$. The coefficients for the best fits obtained in Ref.~\cite{sotani:2021urb} are presented in Table~\ref{table:fitsotaninu} for Eq.~\ref{eq:nudimL}, and in Table~\ref{table:fitsotanitau} for Eq.~\ref{eq:taudimL}. In Fig.~\ref{fig:unilambdanu}, we show that the results for $\nu_f$ corresponding to SSHSs, constructed with hybrid EoS with $\Delta \varepsilon \gtrsim 1000$~MeV/fm$^3$ do not present universal behavior. As in the previous cases, the same occurs for the damping time, but again the deviation is already visible for smaller values of $\Delta \varepsilon$. We present these results in Fig.~\ref{fig:unilambdatau}. The previously mentioned and anomalous $\Delta \varepsilon = 1000$~MeV/fm$^3$ case show, in this particular UR, a remarkable deviation. In general, as in all the previous URs figures, the effects of deviation are more notorious -and began to occur for smaller $\Delta \varepsilon$ values- for the $\tau_f$ relationships, than for the $\nu_f$ ones.




\section{\label{sec:conclus} Summary and Discussion}

In this work, we have calculated the quadrupolar fundamental $f$-mode of compact objects constructed with hybrid EoSs that possess a sharp hadron-quark phase transition. In order to obtain results of great generality that can be interpreted to be model-independent, we have used a \textit{soft} and a \textit{stiff} generalized piecewise polytropic hadronic EoS and a wide range of values for the three parameters of the CSS parametrization for quark matter. We analyzed the impact of assuming that the hadron-quark conversion speed at the interface is slow compared to the characteristic timescale of radial oscillation modes. This situation dramatically alters the dynamic stability of compact objects and gives rise to branches of SSHS.

We have tested all existing proposed URs for both the frequency and the damping time of the $f$-mode and found that SSHS systematically deviate from universality. Besides this global behavior, we found that the deviations from URs are more notorious for greater values of the hadron-quark energy jump $\Delta \varepsilon$; in particular, most of the deviations come from the EoSs with $\Delta \varepsilon \gtrsim 1000$~MeV/fm$^3$. {Generally speaking, the existence of long SSHS branches spanning to low values of masses is key to the breaking of every UR. This is related to the value of $\Delta \varepsilon$ as larger values of this CSS parameter lead to longer SSHS branches of stellar objects. This association between large values of $\Delta \varepsilon$ and the length of the SSHS branch of objects has recently been studied in \cite{Lenzi:2023hsw}.} 

{Moreover, we have seen that the} variation of the speed of sound $c_\mathrm{s}$ also induces noticeable changes, both in the macroscopic quantities -such as mass, radius, length of the SSHS branch, tidal deformability, and moment of inertia- and in the QNM quantities; nevertheless, there exist a bigger dependence of all these quantities to the energy jump $\Delta \varepsilon$, and this behavior also applies to the URs breaking. On the other hand, the breaking is more noticeable for the $\tau_f$ URs than for the $\nu_f$ cases; in particular for the $\Lambda$ and $I$ URs, and even for smaller values of $\Delta\varepsilon \sim 500$~MeV/fm$^3$.

With a broader perspective, within the SSHS hypothesis, still to be elucidated, these results pose some concerns related to the astronomical applicability of these known asteroseismology tools. These results are in concordance with those obtained for $wI$-modes in Ref.~\cite{RaneaSandoval:2022bou}, where it has been shown that for SSHS, both frequency and damping time also shows a clear deviation from \textit{universal} behavior. Regarding these $wI$-modes, in Ref.~\cite{ranea-sandoval:2023cmr}, URs for wI-modes that include SSHSs have been presented, and some of their astronomical applications were shown. Although it would represent a key development, the construction of such \textit{new} URs for the $f$ mode is beyond the scope of the present work.

It should be pointed out that SSHS also exist for the case when the phase transition takes place in stellar configurations of lower mass (see e.g. Ref. \citep{Pereira_2021}), however, in this case, the deviation from the URs presented in this work is much smaller. The strong deviations occur due to a combination of transition pressure and energy density jump that render possible the existence of extreme {low-mass} objects in the stable extended branch. {Although we might expect some effect on the URs due to the inclusion of these extreme objects, our objective here is to highlight that the standard URs do not take into account the possibility of SSHS, which may still exist.}

{The 1-node behavior of the $f$-mode shown in Ref.~\cite{zhao:2022urf} seems to be general in the case of the SSHS studied in this work. Despite this fact, in these cases, the perturbation functions can not be classified either as 1-node I or 1-node II{, according to the definition of the mentioned work}. Moreover, the relationship between this {novel} behavior of the $f$-mode and the violation of URs that was pointed out in Ref.~\cite{zhao:2022urf} can not be {confirmed nor} discarded from our results.} {In this sense, our results would suggest that while the studied SSHSs present the 1-node behavior, only the low-mass SSHS configurations belonging to long enough extended stability branches contribute to the violation.} {A deeper analysis that lies out of the scope of this work is needed to draw more conclusive statements regarding this fact.}

As we had introduced, the $f$-modes modes might be detected by the next-generation gravitational-wave observatories, like the Einstein Telescope (see, for example, Refs.~\cite{Punturo:2010zz,maggiore:2020scf} and references therein) or the Cosmic Explorer (see, for example, Ref.~\cite{Hall:2021gwp}, and references therein). Moreover, a kilohertz-band gravitational-wave detector, the Neutron star Extreme Matter Observatory (NEMO) has been proposed \cite{NEMO:2020}. If it is ever built, joint observations of this detector and the LIGO-Virgo Collaboration are expected to be key to shedding some light on the nature of the matter in the inner core of extremely compact stars and on the existence of the SSHS.

\begin{acknowledgments}
The authors thank the anonymous referee for the constructive comments and criticisms that have contributed to improve the manuscript substantially. I.F.R-S, M.M, M.O.C and M.C.R acknowledge CONICET and UNLP for financial support under grants G157 and G007. M.M is a postdoctoral fellow of CONICET. M.O.C and M.C.R are doctoral fellows of CONICET.  I.F.R-S is also partially supported by PICT 2019-0366 from ANPCyT (Argentina) and by the National Science Foundation (USA) under Grant PHY-2012152. L.T thanks the Italian Istituto Nazionale di Fisica Nucleare (INFN) under grant TEONGRAV.

\end{acknowledgments}

\appendix

\section{\label{app:tables} URs coefficient values}

In this appendix, we present the coefficient values for all of the URs from the literature that we study in this work. In order to obtain these numerical values, the mass $M$ and radius $R$ should be in km, the moment of inertia $I$, in km$^3$, the frequency $\nu_f$ in kHz and the damping time $\tau_f$, in seconds. We have convert the values for every coefficient in order that they become consistent with our selection of units.

\begin{table}[h!]
\begin{tabular}{lcc}
\toprule
Fit & $a_\nu$ & $b_\nu$ \\
\midrule
AK \cite{Anderson:1998tgw} & 0.78 & 35.919 \\
BFG \cite{Benhar:2004gwa} & 0.79 $\pm$ 0.09 & 33 $\pm$ 2  \\
CSK \cite{chirenti:2015PhRvD} &  0.332 $\pm$ 0.275 & 44.04  \\
\bottomrule
\end{tabular}
\caption{Parameter values for Eq.~\ref{eq:freqfit} for fits in different previous works.
}
\label{table:fitsnu}
\end{table}

\begin{table}[h!]
\begin{tabular}{lcc}
\toprule
Fit & $a_\tau$ & $b_\tau$ \\
\midrule
AK \cite{Anderson:1998tgw} & 25687  & -79483 \\
BFG \cite{Benhar:2004gwa} & 26100 $\pm$ 600 & -81300 $\pm$ 2700  \\
CSK \cite{chirenti:2015PhRvD} & 25200 $\pm$ 3600 & -78000  \\
\bottomrule
\end{tabular}
\caption{Parameter values for Eq.~\ref{eq:taufit} for fits in different previous works.
}
\label{table:fitstau}
\end{table}

\begin{table}[h!]
\begin{tabular}{lccc}
\toprule
Fit & $a_\nu^I$ & $b_\nu^I$ & $c_\nu^I$ \\
\midrule
CSK \cite{chirenti:2015PhRvD} & -7.214 $\pm$ 4.144 & 240.52 & 1077.82 \\
LLL \cite{lau:2010ipp} & -8.859 & 250.699 & 1083.85 \\
\bottomrule
\end{tabular}
\caption{Parameter values for Eq.~\ref{eq:inertianu} for fits in different previous works.
}
\label{table:inertianu}
\end{table}

\begin{table}[h!]
\begin{tabular}{lcc}
\toprule
Fit & $a_\tau^I$ & $b_\tau^I$ \\
\midrule
CSK \cite{chirenti:2015PhRvD} & 2040 $\pm$ 39 & -7500 \\
LLL \cite{lau:2010ipp} & 2082 & -7680 \\
\bottomrule
\end{tabular}
\caption{Parameter values for Eq.~\ref{eq:inertiatau} for fits in different previous works.
}
\label{table:inertiatau}
\end{table}

\begin{table}[h!]
\begin{tabular}{lcccccc}
\toprule
Fit & $a^\Lambda_\nu$ & $b^\Lambda_\nu$ & $c^\Lambda_\nu$ & $d^\Lambda_\nu$ & $e^\Lambda_\nu$ & $f^\Lambda_\nu$ \\
\midrule 
SK \citep{sotani:2021urb} & 8.8246 & -0.9919 & -0.9397 & 0.2991 & -0.0335 & 0.0013 \\
\bottomrule
\end{tabular}
\caption{Parameter values for Eq.~\ref{eq:nudimL} for the fit presented in \cite{sotani:2021urb}.
}
\label{table:fitsotaninu}
\end{table}

\begin{table}[h!]
\begin{tabular}{lcccccc}
\toprule
Fit & $a^\Lambda_\tau$ & $b^\Lambda_\tau$ & $c^\Lambda_\tau$ & $d^\Lambda_\tau$ & $e^\Lambda_\tau$ & $f^\Lambda_\tau$ \\
\midrule 
SK \citep{sotani:2021urb} & 1.1433 & 0.4589 & -0.2798 & 0.03651 & 0.0025 & 6.2574$\times10^{-5}$ \\
\bottomrule
\end{tabular}
\caption{Parameter values for Eq.~\ref{eq:taudimL} for the fit presented in \cite{sotani:2021urb}.
}
\label{table:fitsotanitau}
\end{table}


\renewcommand{\href}[2]{#2}

\bibliography{biblio}

\end{document}